\documentclass[useAMS,usenatbib]{mn2e}
\usepackage{amsmath,graphicx,txfonts}
\newcommand{\mr}{\mathrm}

\newcommand{\be}{\begin{equation}}
\newcommand{\ee}{\end{equation}}

\newcommand{\beq}{\begin{eqnarray}}
\newcommand{\eeq}{\end{eqnarray}}

\def\lsim{\mathrel{\rlap{\lower4pt\hbox{\hskip1pt$\sim$}}
    \raise1pt\hbox{$<$}}}                
\def\gsim{\mathrel{\rlap{\lower4pt\hbox{\hskip1pt$\sim$}}
    \raise1pt\hbox{$>$}}}                
\def\aap{A\&A}
\def\apj{ApJ}
\def\apjl{ApJL}
\def\apjs{ApJS}
\def\mnras{MNRAS}
\def\aj{AJ}
\def\nat{Nature}

\def\prd{Phys.~Rev.~D}        

\begin{document}
\title[Merger induced scatter in the $M$-$Y$ relation]{Merger induced scatter and bias in the cluster mass - Sunyaev-Zeldovich effect scaling relation}
\author[E. Krause et al.]{Elisabeth Krause$^1$, Elena Pierpaoli$^2$, Klaus Dolag$^{3,4}$, Stefano Borgani$^5$ 
\\$^1$California Institute of Technology, MC 249-17, Pasadena, CA, 91125, USA, ekrause@tapir.caltech.edu
\\$^2$University of Southern California, Los Angeles, CA, 90089-0484, USA
\\$^3$Max-Planck Institut f\"ur Astrophysik, Karl-Schwarzschild Str.1, D-85748 Garching, Germany 
\\$^4$University Observatory Munich, Scheinerstr. 1, D-81679 Munich, Germany
\\$^5$Astronomy Unit, Department of Physics, University of Trieste, via Tiepolo 11, I-34131 Trieste, Italy
}
\date{}
\pagerange{\pageref{firstpage}--\pageref{lastpage}} \pubyear{2011}
\maketitle
\label{firstpage}
\begin{abstract}
We examine sources of scatter in scaling relations between galaxy cluster mass and thermal Sunyaev-Zeldovich (SZ) effect using cluster samples extracted from cosmological hydrodynamical simulations. Overall, the scatter of the mass-SZ scaling relation is well correlated with the scatter in the mass-halo concentration relation with more concentrated halos having stronger integrated SZ signals at fixed mass. Additional sources of intrinsic scatter are projection effects from correlated structures, which cause the distribution of scatter to deviate from log-normality and skew it towards higher inferred masses, and the dynamical state of clusters. We study the evolution of merging clusters based on simulations of 39 clusters and their cosmological environment with high time resolution. This sample enables us to study for the first time the detailed evolution of merging clusters around the scaling relation for a cosmologically representative distribution of merger parameters. Major mergers cause an asymmetric scatter such that the inferred mass of merging systems is biased low. We find mergers to be the dominant source of bias towards low inferred masses: over 50\% of outliers on this side of the scaling relation underwent a major merger within the last Gigayear. As the fraction of dynamically disturbed clusters increases with redshift, our analysis indicates that mergers cause a redshift-dependent bias in scaling relations. Furthermore, we find the SZ morphology of massive clusters to be well correlated with the clusters' dynamical state, suggesting that morphology may be used to constrain merger fractions and identify merger-induced outliers of the scaling relation.
\end{abstract}
\begin{keywords}
cosmology: large-scale structure of Universe, galaxies: clusters: general, methods:  N-body simulations\end{keywords}
\section{Introduction}
Clusters of galaxies are the most massive gravitationally bound objects in the universe, which makes them an important tool for cosmology: among other tests, their abundance provides information on the gravitational growth of structures and is regulated by the initial density field, gravity, and the expansion history of the universe, which critically depend on the underlying cosmology. Thus number counts of clusters, for which masses and redshifts are known, can be used to constrain cosmological parameters \citep[see][for a recent review]{Allen11}.\\
To relate observed number counts to theoretical predictions of the cluster mass function, these experiments need to infer cluster masses from observables. The thermal Sunyaev Zeldovich (SZ) effect, the signature of inverse Compton scattering of cosmic microwave background photons with hot cluster electron, is thought to provide an excellent mass proxy as the SZ signal is proportional to the total thermal energy of a cluster and is thus less affected by physical processes in the cluster core which can largely affect the X-ray luminosity. This is confirmed by simulations \citep[e.g.][]{Nagai06, Shaw08, Battaglia10, Sehgal10} finding the scatter in the mass - SZ scaling relation to be of order 5 - 10\%. Furthermore, the SZ effect is not subject to surface brightness dimming and has a very weak redshift dependence, making it an ideal probe to study galaxy clusters at high redshift.\\

Currently several large surveys are starting to detect hundreds of galaxy clusters through their SZ signal \citep{Vanderlinde10,Marriage10, Planck11} and derive cosmological constraint based on these samples \citep{Andersson10, Sehgal10b, Williamson11}. To exploit the statistical power of these upcoming cluster samples, the mapping between SZ signal and cluster mass needs to be well understood. Observations find normalization and slope of the scaling relations between SZ signal and lensing derived masses \citep{Marrone11}, or between SZ signal and X-ray properties \citep{Planck11b, Planck11c} to be consistent with self-similar scaling and predictions from simulations.\\

Due to the steep slope of the cluster mass function, competitive cosmological constraints from these experiments require information about the distribution and redshift evolution of scatter in the mass scaling relation \citep[e.g.][]{Majumdar04, Lima05,Shaw10}. As the true cluster mass and other physical cluster properties which may bias the mass proxy are unobservable, and as the noise and biases in the different mass estimators may be correlated, characterizing the intrinsic scatter in any of these scaling relation is difficult to obtain from observations. Hence the sources and distribution of scatter in different mass estimators are mainly studied through simulations and mock observations \citep[e.g.][]{Rasia06,Nagai07,Shaw08,Becker10,Yang10,Fabjan11}.\\

In this work we focus on the effect of merging events on the SZ signal of a galaxy cluster. As clusters form through merging of smaller objects these are frequent and disruptive events, which may alter the physical state of the involved clusters significantly. Hence merging clusters may deviate from the scaling relations observed in relaxed clusters and, as the fraction of morphologically disturbed clusters increases with redshift, cause a redshift dependent scatter or bias in the mass scaling relation. Simulations of binary cluster mergers \citep{Randall02,Poole06, Poole07, Wik08} find that the X-ray luminosities, temperatures, SZ central Compton parameters and integrated SZ fluxes increase rapidly during the first and second passage of the merging clusters. The clusters temporarily drift away from mass scaling relations and return to their initial scaling relation as the merging system virializes. These transient merger boosts found in binary mergers and some observations \citep{Smith03} can scatter the inferred masses towards higher values and thus bias the derived cosmology towards a higher normalization of the power spectrum, $\sigma_8$, and lower matter density \citep{Randall02,Smith03,Wik08,Angrick11}. On the other hand, mergers increase the non-thermal pressure support \citep{Rasia06, Lau09, Battaglia10} found in cluster outskirts, and due to partial virialization merging clusters can appear cooler than relaxed clusters of the same mass \citep[e.g.][]{Mathiesen01}. For a cluster sample extracted from cosmological simulations, \citet{Kravtsov06} find the X-ray temperatures of morphologically disturbed clusters to be biased, while the X-ray derived SZ-equivalent $Y_X$ shows no significant correlation with cluster structure. Comparing X-ray and SZ to weak lensing derived masses, \citet{Okabe10} and \citet{Marrone11} found undisturbed clusters to have of order $\sim\!40\%$ higher weak lensing masses than disturbed clusters at fixed $T$ and $Y_{\mr{SZ}}$, and $\sim\!20\%$ higher weak lensing masses at fixed $Y_{\rm X}$.\\

Our goal is to isolate how mergers in a cosmological context affect the SZ signal of clusters, and if merging cluster can be detected as outliers of scaling relations. This extends previous work, as our analysis includes both multiple mergers with realistic distributions of orbits and mass ratios, and full SPH treatment of gas physics with radiative cooling, star formation and supernova feedback. The simulations and the cluster sample are described in Sect. \ref{sec:simul}. We discussion the best-fit scaling relations and their scatter in Sect. \ref{sec:scaling}. The effect of merging events of the clusters SZ signal is quantified and the evolution of merging clusters with respect to the scaling relations is discussed in Sect. \ref{sec:merger}. In Sect. \ref{sec:morph} we investigate if the dynamical state of clusters can be inferred from the morphology of the SZ signal. We summarize our results and conclude in Sect.~\ref{sec:summary}.
\section{Simulations}
\label{sec:simul}
This analysis is based on two samples of galaxy clusters extracted from cosmological hydrodynamics simulations. In this section we summarize the simulated physics and describe the derived quantities used in our analysis.
\subsection{Cluster samples}
\paragraph*{Sample A}
To study the time evolution of the cluster SZ signal we use a sample of 39 galaxy groups and clusters with virial masses above $3\times 10^{13} M_\odot/h$ from simulations presented in \citet{Dolag06, Dolag09}. 25 of these clusters are more massive than $10^{14} M_{\odot}/h$. These structures were identified as 10 different regions in a $(479 \mr{Mpc}/h)^3$ dark-matter-only cosmological simulation \citep{Yoshida01}, and re-simulated at higher resolution using the Zoomed Initial Conditions method \citep{Tormen97}. The re-simulations, described in detail in \citet{Dolag06}, are carried out with GADGET-2 \citep{Springel05}, and include a uniform, evolving UV-background and radiative cooling assuming an optically thin gas of primordial composition. Star formation is included using the two-phase model of the interstellar medium (ISM) by \citet{Springel03}. In this sub-resolution model the ISM is described as cold clouds, providing a reservoir for star formation, embedded in the hot phase of the ISM. Star formation is self-regulated through energy injection from supernovae evaporating the cold phase. Additional feedback is incorporated in the form of galactic winds triggered by supernovae that drive mass outflows \citep{Springel03}. \\
The simulation assumes a flat $\Lambda$CDM cosmology with $(\Omega_{\mr m},\Omega_{\mr b}, \sigma_8, h) =(0.3,0.04,0.9,0.7)$. It has a mass resolution of $m_{\rm{DM}} = 1.1\times 10^9 M_\odot/h$ and $m_{\rm{gas}} = 1.7\times 10^8 M_\odot/h$ and the physical softening length is $\epsilon = 5\mr{kpc}/h$ over the redshift range of interest. Our analysis is based on 52 snapshots covering the redshift range $z=1$ to $z=0$ and separated evenly in time with a spacing of 154 Myrs between snapshots.\\

\paragraph*{Sample B}
The second cluster sample is a volume-limited sample of 117 clusters at $z=0$ described in \citet{Borgani04}. These clusters are identified in a $(192 \mr{Mpc}/h)^3$ cosmological SPH simulation carried out with GADGET-2 and using the same physics as described above. This simulation assumes a flat $\Lambda$CDM cosmology with $(\Omega_{\mr m},\Omega_{\mr b}, \sigma_8, h) =(0.3,0.04,0.8,0.7)$. The mass resolution is $m_{\rm{DM}} = 4.6\times 10^9 M_\odot/h$ and $m_{\rm{gas}} = 6.9\times 10^8 M_\odot/h$, the physical softening length at $z=0$ is $\epsilon = 7.5\mr{kpc}/h$.

\subsection{Masses and merging histories}
\begin{figure}
\includegraphics[width = 0.5\textwidth, trim  = 0mm 0mm 0mm 0mm, clip = true]{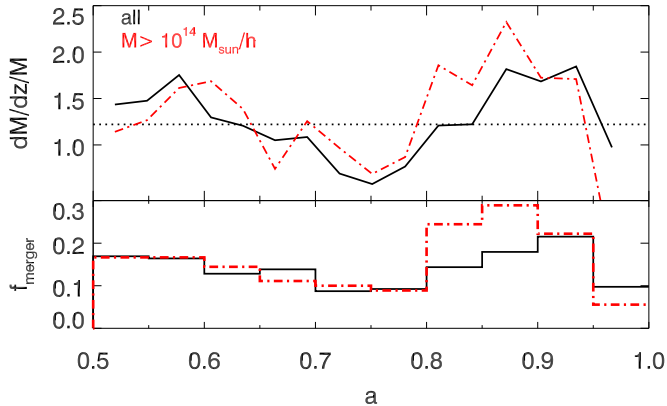}
\caption{Fractions accretion rate per unit redshift (top panel) and merger fraction as a function of scale factor. The solid shows the complete sample A, the dash-dotted line a subsample of massive clusters. The dotted line indicates the overall mean accretion rate. Accretion rate (merger fraction) are averaged over 3 (5) neighboring simulation snapshots to reduce noise.}
\label{fig:dmdz}
\end{figure}
Halos are identified using a friend-of-friends algorithm and the cluster center is defined by the particle in a halo with the minimum gravitational potential. Cluster radii $R_\Delta$ and masses $M_\Delta$ are defined through spherical regions around the cluster center within which the average density is $\Delta$ times the critical density of the universe
\be
\int_0^{R_{\Delta}} \rho (r)\ 4\pi r^2 \ dr = \frac{4 \pi}{3}R_{\Delta}^3 \Delta\, \rho_{\rm{crit}} = M_\Delta\ \  .
\ee

We identify mergers by a mass jump criteria applied to the mass history of the main progenitor. Motivated by the findings that the average mass accretion history of halos is well described by exponential growth with redshift \citep{Wechsler02, McBride09} and that the average merger rate per halo per unit redshift is nearly constant for a wide range of halo masses and redshifts \citep{Fakhouri08}, we select merging events based on a threshold in fractional mass accretion rate per unit redshift $\mr{d} M/\mr{d}z/M > \zeta_{\mr m}$. We choose $\zeta_{\mr m}$ such that halos accrete on average $30\%$ of the mass accreted since its formation redshift  $z_\mr{f}$, defined as the redshift at which a halo reaches half its present day mass, during mergers. We checked that our results are insensitive to the exact choice of $\zeta_{\mr m}$: We find similar trends for any merger definition $\zeta_{\mr M} \ge \langle \mr{d} M/\mr{d}z/M \rangle_{z,\mr{cluster}}$ that requires the accretion rate $\mr{d} M/\mr{d}z/M$ during mergers to be larger than the mean accretion rate (cf. discussion of Fig.~\ref{fig:evol2}).\\

Figure~\ref{fig:dmdz} confirms that this merger definition does not strongly depend on cluster mass or redshift. The top panel shows the mean accretion rate as a function of scale factor for all clusters (solid line) and massive clusters ($M\ge 10^{14}M_{\odot}/h$, dash dotted line), and the overall mean accretion rate (dotted line). The lower panel shows the fraction of clusters that are merging as a function of scale factor. There is a peak of merging activity around $a =0.9$, but the accretion rate and merger fraction show no clear trends with cluster mass or redshift.

\subsubsection{Comparison to the Millenium Run}
The 39 cluster and group-scale sized halos in sample A are extracted from 10 re-simulation regions selected from a large simulation box. One of the re-simulated regions hosts a filamentary structure with four massive clusters ($M>10^{15} M_\odot/h$), and three of the re-simulation regions hosting other massive clusters contain several other smaller clusters. The re-simulation technique allows us to analyze the evolution of these regions of interest in their cosmological context at a higher resolution. As a result of the re-simulation strategy, the mass distribution of this sample does not follow the cluster mass function, and clusters which are not the most massive object in their re-simulation region live in denser regions than an average cluster of the same mass in a volume limited sample. In the following discussion we refer to the most massive objects in their respective re-simualtion region as primary clusters, and all others as secondary clusters.

\begin{figure}
\includegraphics[width = 0.5\textwidth, trim  = 0mm 0mm 0mm 0mm, clip = true]{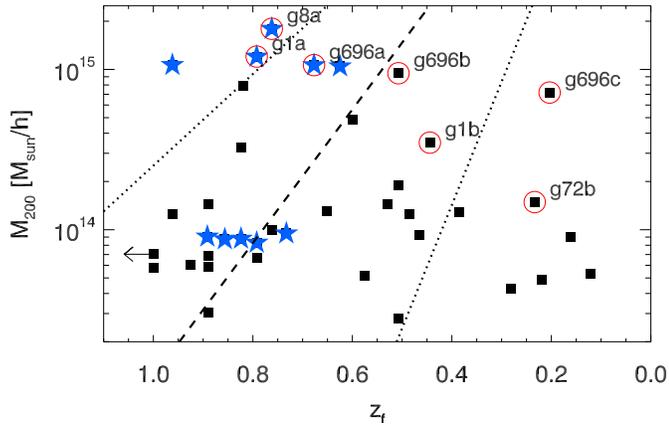}
\caption{Distribution of formation redshifts $z_{\mr f}$ as a function of cluster mass. The symbols represent clusters from sample A, the most massive cluster in each re-simulation region is marked with a star symbol. Dashed (dotted) lines show a fit to the mean ($1\sigma$ scatter) formation redshift as a function of friends-of-friends mass found in the Millenium Run \citep{McBride09}, converted to spherical overdensity mass assuming $M_{200}  = 0.7 M_{\mr{FOF}}$ (see text for details). Formation redshift is defined as the redshift at which a halo reaches half its present day mass. One cluster in sample A forms before $z = 1$, indicated by the left arrow. Open circles indicate the clusters shown as examples in subsequent plots, labels indicate the names of these clusters in Table~1 of \citet{Dolag09}.}
\label{fig:af}
\end{figure}

Simulations indicate a dependence of halo formation histories on environment with merger being more frequent in dense environments and late-forming massive clusters living in denser environments than earlier forming clusters of the same mass \citep{Gao05,Wechsler06, Fakhouri09}. Hence the merging histories of cluster sample A might not be representative of those of a volume limited sample. To assess the impact of our sample selection on halo formation histories we compare the formation redshifts of primary and secondary clusters in sample A and halos in the Millenium run simulation \citep{Millenium} in Fig.~\ref{fig:af}.

The symbols show the present day masses and formation redshift  $z_\mr{f}$ for all clusters in sample A. Primary clusters are indicated by star symbols. The dashed and dotted lines are a fit to the mean formation time and its $1\sigma$ scatter for halos in the Millenium Run from \citet{McBride09}. 
We convert the fitting formula from friend-of-friends halo mass with linking length $b=0.2$ to $M_{200}$ assuming a constant conversion factor $M_{200}  = 0.7 M_{\mr{FOF}}$. For the mass range of our sample this conversion underestimates $M_{200}$
\footnote{For equal mass particles, a FOF group with linking length $b$ is bounded by a surface of density $3\Omega_{\mr{m}}\,\rho_{\mr{crit}}/(2 \pi b^3) $ \citep{White02}. Assuming that halos follow NFW-profiles with concentration $c = (4,7,10)$, the ration between $M_{200}$ and $M_{\mr{FOF}}$ with $b=0.2$ in the Millenium run cosmology is given by $(0.71,0.80,0.85)$. In practice however, the conversion between these mass definitions is complicated by deviations from the NFW-profile and spherical symmetry.} 
and biases the fit for $z_{\mr{f}}$ to more recent times. 

Due to the differences in matter density used in simulation A ($\Omega_{\mr M} = 0.3$) and in the Millenium Run ($\Omega_{\mr m} = 0.25$) the average clusters in simulation A forms earlier than a cluster of the same mass in the Millenium Run.
Hence formation redshifts for primary clusters in sample A are broadly consistent with the formation history of halos in the Millenium run. Figure~\ref{fig:af} suggests that secondary clusters in sample A may form somewhat later than primary clusters of the same mass. However, the distribution of formation redshifts at fixed mass is not expected to be symmetric but to have a long tail towards later formation times and the comparison is limited by the small number objects. Overall, we expect the merging histories analyzed in this study to be similar to those found in a volume limited sample.

\subsection{SZ maps}
The amplitude of the thermal SZ effect along a line of sight is proportional to the Compton $y$ parameter
\be
y =  \frac{k_{\mr B}\sigma_{\mr T}}{m_{\mr e}c^2} \int \mr d l\, n_{\mr e} T_{\mr e}\,,
\ee
where $n_{\mr e}$ and $T_{\mr e}$ are the electron density and temperature, $k_{\mr B}$ is the Boltzmann constant, $\sigma_{\mr T}$ the Thomson cross section, $m_{\mr e}$ the electron rest mass, and $c$ the speed of light.
For each cluster we analyze Compton $y$ parameter maps obtained from three orthogonal lines of sight. For sample A the projection depth is 8 Mpc and maps are produced using the map making tool Smac \citep{Dolag05} and the JobRunner web application\footnote{
Access to the cluster simulations of sample {\it A}, including web services allowing
to interactively produce various kinds of maps, are publicly available via 
the web portal at http://www.mpa-garching.mpg.de/HydroSims}. For sample B we use projected maps which include all material with $6 R_{\mr{vir}}$ described in \citet{Ameglio07}. From these maps we measure integrated $Y_\Delta$ parameters within different overdensity radii ($R_{2500},R_{500},R_{200},R_{\rm{vir}}$)
\be
Y_\Delta = \frac{k_{\mr B}\sigma_{\mr T}}{m_{\mr e}c^2}\int_{V_{\Delta}} \mr d V\,  n_{\mr e} T_{\mr e} 
\ee
where the integration volume is a cylinder of radius $R_\Delta$ and height 8 Mpc (or 12 $R_{\mr{vir}}$) for sample A (or B). This definition of the integrated $Y$ parameter includes projection effects due to halo triaxiality and nearby structures within the projection cylinder, but does not account for projection effects from uncorrelated large scale structure along the line of sight.

\section{Mass scaling relations}
\label{sec:scaling}
Self-similar clusters models predict the gas temperature to scale as
\be
T \propto \left(M E(z)\right)^{2/3}\,.
\ee
Hence the self-similar prediction for the relation between integrated Compton $Y$ parameter and mass is 
\be
Y_{\Delta} \propto  M_{\mr{gas},\Delta} T \propto f_{\mr {gas}}M_{\Delta}^{5/3}E^{2/3}(z)\,.
\ee
In this section we determine the best fit scaling relations for the simulated clusters and discuss the scatter in these relations, focussing on the role of mergers.
\subsection{Best fit scaling relations}
\begin{table}
\caption{Best fit $M_\Delta(Y_\Delta)$ scaling relation parameters  (Eq.~(\ref{eq:my}))and logarithmic scatter $\sigma_M$ at fixed $Y$, defined analogously to Eq.~\ref{eq:sigmam}. A*/B* denote samples A/B restricted to clusters at $z= 0$ with $M > 2\times 10^{14} M_\odot/h$. }
\begin{center}
\begin{tabular}{l c || c c c |c}
\hline
\hline
Sample & $\Delta$ & $A (z = 0)$ & $\alpha$ & $\beta$&$ \sigma_M$\\
\hline
\hline
A & 200 &$ -0.348\pm0.007$& $0.639\pm0.010$& $-0.57\pm0.08$&0.063\\
A*  & 200 &$-0.281\pm 0.042$ &$0.588\pm0.020$ & - &0.042\\
B & 200 & $-0.297\pm0.006$&$ 0.617\pm 0.007$& - & 0.042\\
B*  & 200 &$-0.261\pm 0.014$ &$0.593\pm0.010$ & - &0.027\\
\hline
A & 500 & $-0.466\pm 0.001$& $0.641\pm 0.007$& $-0.74\pm 0.10$&0.089\\
A* & 500 & $-0.406\pm 0.036$&$0.607\pm0.020$ & - &0.042\\
B & 500 & $-0.400\pm0.004$& $0.626\pm 0.005$& - &0.037\\
B* & 500 & $-0.379\pm 0.011$&$0.604\pm0.009$ & - &0.024\\
\hline
\hline
\end{tabular}
\end{center}
\label{tab:my}
\end{table}
\begin{table}
\caption{Best fit $Y_\Delta(M_\Delta)$ scaling relation parameters (Eq.~(\ref{eq:ym})) and logarithmic scatter $\sigma_Y$ at fixed mass. A*/B* denote sample A/B restricted to clusters at $z =0$ with $M > 2\times 10^{14} M_\odot/h$. }
\begin{center}
\begin{tabular}{c c || c c c |c}
\hline
\hline
Sample & $\Delta$ & $B (z = 0)$ & $\gamma$ & $\delta$&$ \sigma_Y$\\
\hline
\hline
A & 200 & $0.547\pm 0.003$& $1.560\pm 0.014$&$0.85\pm0.10$&0.103\\
A*  & 200 &$0.489\pm 0.052$ &$1.648\pm0.056$ & - &0.070\\
B & 200 & $0.494\pm0.005$&$ 1.555\pm 0.017$& - & 0.071\\
B*  & 200 &$0.445\pm 0.030$ &$1.668\pm0.044$ & - &0.046\\
\hline
A & 500 & $0.714\pm 0.003$& $1.553\pm 0.017$& $1.03\pm 0.14$&0.136\\
A*  & 500 &$0.697\pm 0.038$ &$1.601\pm0.051$ & - &0.068\\
B & 500 & $0.641\pm0.003$& $1.556\pm 0.014$& - &0.059\\
B* & 500 & $0.624\pm 0.013$&$1.637\pm0.027$ & - &0.037\\
\hline
\hline
\end{tabular}
\end{center}
\label{tab:ym}
\end{table}
We now determine the best fit $M_\Delta(Y_\Delta)$ scaling relation
\be
M_\Delta(Y_\Delta) = 10^A \left(\frac{Y_\Delta}{\mr{kpc}^2}\right)^\alpha E^\beta (z)\, 10^{14} M_{\odot}/h
\label{eq:my}
\ee
and $Y_\Delta(D_\Delta)$ scaling relation
\be
Y_\Delta(M_\Delta) = 10^B \left(\frac{M_\Delta}{10^{14} M_{\odot}/h}\right)^\gamma E^\delta (z)\, \mr{kpc}^2\,,
\label{eq:ym}
\ee
where the self similar predictions are $(\alpha,\beta) = (3/5,-2/5)$ and $(\gamma,\delta) = (5/3,2/3)$. Specifically we first fit a line to the $\lg(Y_\Delta)-\lg(M_\Delta)$ distribution at each redshift, and then determine the redshift dependence by determining a linear fit in $\lg(E(z))$ to the evolution of the normalization constant $B(z)$. We find no significant indication for a redshift evolution of the slope $\alpha$ or $\gamma$.\\
The best fit parameters and the logarithmic scatter at fixed mass,
\be
\sigma_Y = \left(\frac{\sum_{i=1}^N \left(\lg(Y_i/Y(M_i))\right)^2}{N-2}\right)^{1/2}\,,
\label{eq:sigmam}
\ee
where the sum runs over all $Y$ measurements (three projections of each cluster at each redshift), are given in Table~\ref{tab:my} and Table~\ref{tab:ym}.\\ The two scaling relations contain the same information. While the $M(Y)$ scaling relation is the relation of more interest for cosmology and is the relation used in the rest of our analysis, the $Y(M)$ relation is easier to interpret if one is more used to thinking about clusters properties at fixed mass rather than at fixed $Y$, and we will focus the discussion of the fit results on this relation.\\

The slope $\gamma$ of the best fit relation in samples A and B is below the self similar value while other simulations including cooling and star formation find slopes comparable to or steeper than the self similar predictions \citep{Nagai06,Battaglia10,Sehgal10}. We find a slope in agreement with previous results if we only consider massive clusters with $M_{200} > 2\times 10^{14} M_\odot /h$ (``Sample B*'') which is identical to the mass threshold used in \citet{Sehgal10}. Projection effects may account for some of the difference with the results of \citet{Nagai06} and \citet{Battaglia10}: these authors use spherically averaged $Y$ measurements and do not include projection effects, which effectively boost the integrated $Y$ signal of lower mass clusters\footnote{Projection effects introduce an additive signal $Y_{\mr p} \ge0$ which scales as $Y_{\mr p,\Delta} \propto R_{\Delta}^2 \propto M^{2/3}$, and thus the fractional error induced by projection effects decreases with cluster mass} and hence lower the slope of the scaling relation.\\
After accounting for differences in the baryon fractions of different simulations, the normalization $B$ of the best fit scaling relation for sample B* is consistent with those obtained from other hydrodynamical simulations with similar physics (the csf run in \citet{Nagai06} and the radiative run in \citet{Battaglia10}).\\

The slope and normalization of the scaling relation for a subsample of massive clusters at $z=0$ from sample A, denoted as A*, are comparable to those found for the sample B*. A direct comparison of these numbers is complicated by the fact that slope and scatter of the scaling relations are mass dependent, and that the mass distribution within sample A does not follow the cluster mass function. Also sample A* consists of only 11 clusters, five of these are the most massive objects in their respective re-simulation region, and it is hard to assess at a precision cosmology level whether the non-representative environment of clusters in sample A affects the normalization of their scaling relation.\\
The redshift evolution of the scaling relation for sample A deviates significantly from self similar expectations. This deviation may be caused by mergers: As we will discuss in detail in Sect.~\ref{sec:merger} the $Y$ signal of recently merged clusters is suppressed on timescales of order a few Myr. As the merger rate per halo per unit time increases with redshift, the increasing fraction of recently merged clusters  reduces the normalization of the scaling relation, causing $\delta$ to deviate from the self similar value.\\

In the following we will focus on scaling relations within $R_{200}$ as the $M_{200}-Y_{200}$ relation for sample A has less scatter than that within $R_{500}$. The accretion histories at $R_{500}$ are more erratic than at $R_{200}$ which complicates the identification of  merging events and the interpretation of trajectories in the $M-Y$ plane. At the time resolution of the simulation snapshots infalling substructures sometimes cross in and out of $R_{500}$ before coalescence, causing a series of mass jumps and mass losses in $M_{500}$. While it is not clear what the best mass definition is for a merging cluster, the scatter in the $M_{\mr{vir}} -M_{\Delta}$ relation illustrates that masses within larger radii are less volatile: fitting $M_{\Delta}$ as a power law in $M_{\mr{vir}}$ and $E(z)$ we find logarithmic scatter $(\sigma_{M_{200}},\sigma_{M_{500}},\sigma_{M_{2500}})= (0.046, 0.108, 0.326)$.\\

\begin{figure*}
\includegraphics[width = 0.95\textwidth, trim  = 0mm 0mm 0mm 0mm, clip = true]{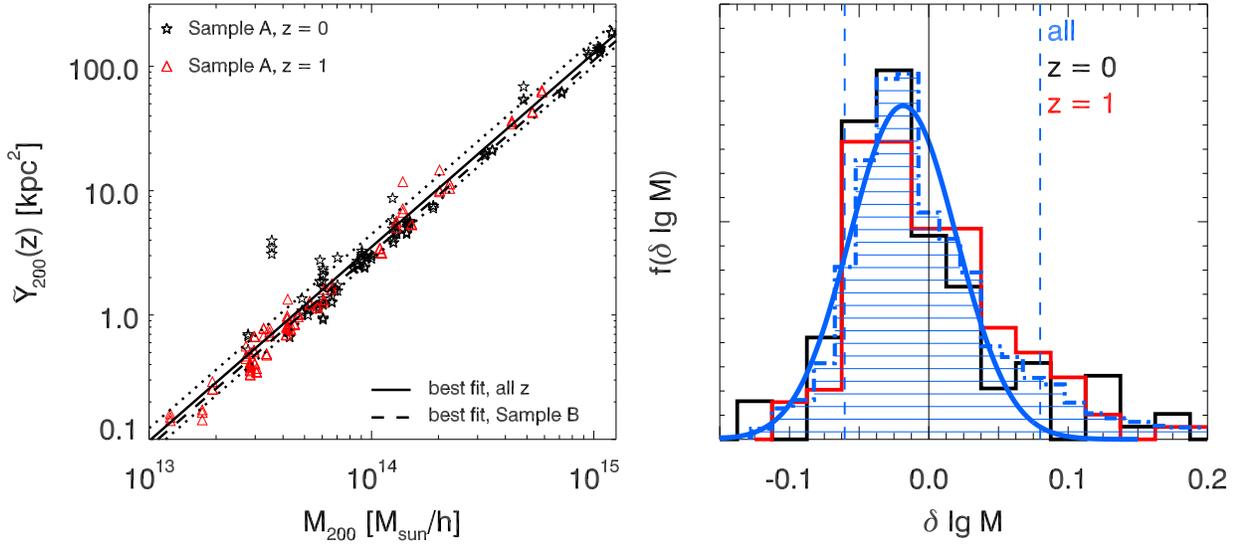}
\caption{
Left: Relation between mass $M_{200}$ and integrated Compton $Y_{200}$ parameter for the $z= 0$ (stars) and $z=1$ clusters (triangles) in sample A. The Compton $Y$ parameter has been scaled to absorb the redshift evolution of the scaling relation in order to show the power law relation $M\propto \tilde Y^\alpha(z) = (Y E^{\beta/\alpha}(z))^\alpha$. The solid and dotted lines show the best fit scaling relation for sample A and its $1\sigma$ error. For reference, the dashed line indicates the best fit scaling relation for sample B.\newline
Right: Distribution of residuals of the best fit scaling relation for the full sample (filled histogram) and the redshift subsamples (black/red line), and the best-fit Gaussian to the full distribution. The vertical dashed lines illustrate the $10\%$ and $90\%$ quantile for the full sample, illustrating the non lognormality of the scatter distribution.}
\label{fig:yM}
\end{figure*}

Figure~\ref{fig:yM} shows the best fit $Y_{200}$-$M_{200}$ scaling relation for sample A and the distribution of the $z=1$ and $z=0$ clusters, which we plot in the form of the SZ signal scaled for redshift evolution
\be
\tilde Y_{200}(z)  =  Y_{200}(z) E^{\beta/\alpha}(z)\,.
\ee
The right panel shows the distribution of the scatter around the scaling relation,
\be
\delta \lg M \equiv \lg \left(M(Y)/M\right)\,,
\ee
for the full sample and subsamples. This scatter definition gives the logarithmic error in the mass inferred from $Y$ measurements, positive scatter corresponds to clusters with $Y$ larger than expected for their actual mass. At all redshifts the distribution deviates from lognormality with a tail at large $\delta \lg M$, causing the distribution to have positive skewness and kurtosis.\\ 
\begin{figure*}
\includegraphics[width = 0.95\textwidth, trim  = 0mm 0mm 0mm 0mm, clip = true]{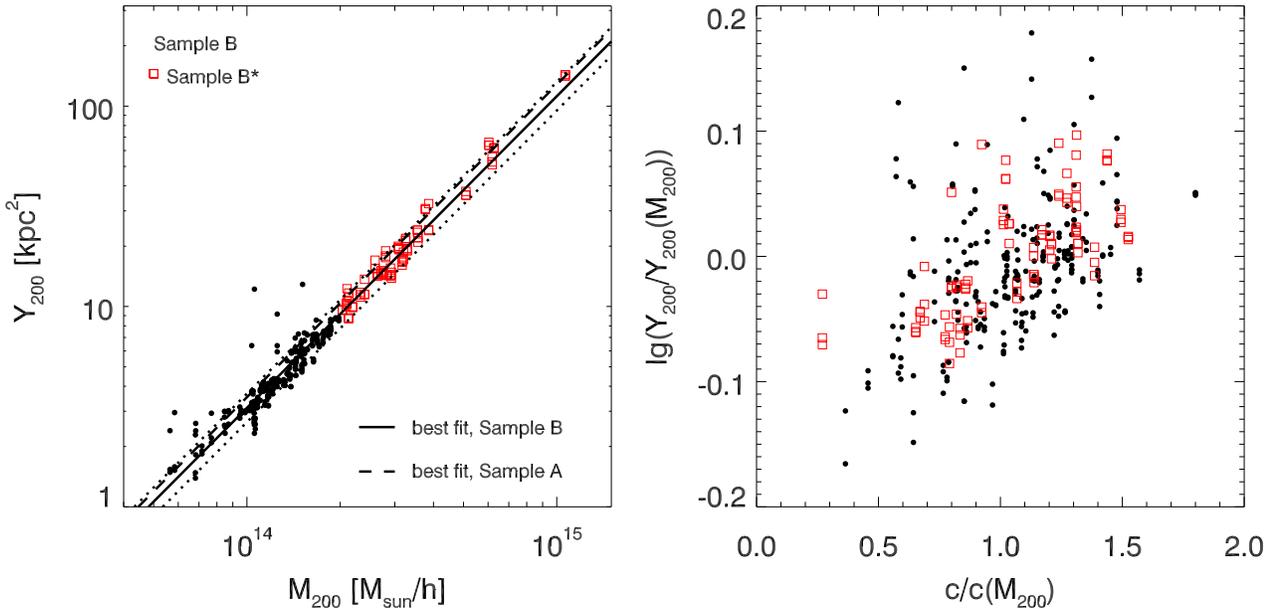}
\caption{Left: Relation between mass $M_{200}$ and integrated Compton $Y_{200}$ parameter for cluster sample B. Massive clusters with $M_{200}> 2\times 10^{14}M_{\odot}/h$ are shown with filled symbols. The solid and dotted lines show the best fit scaling relation for sample B and its $1\sigma$ error. For reference, the dashed line indicates the best fit scaling relation for sample A. The strong outliers with boosted $Y$ signal in the low-mass range are visually identified to be caused by projection effects. \newline
Right: Residuals of the $Y$--$M$ relation at fixed mass vs. scatter in the mass -- halo concentration relation at fixed mass. Concentration measurements are from \citet{Ameglio09}, see text for details on the determination of $c/c(M_{200})$.}
\label{fig:yMz0}
\end{figure*}
The left panel of Fig.~\ref{fig:yMz0} shows the $M_{200}$ and $Y_{200}$ data from sample B and the best fit scaling relation. We checked by visual inspection that the most extreme outliers, which are all in the direction of $Y$ higher than expected for the cluster mass, are indeed projection effects. These clusters have multiple peaks or appear otherwise distorted in only one or two of the three orthogonal projections, indicating that these are not merging systems (yet).\\

The intrinsic scatter in the spherically integrated $Y$ parameter of large cluster samples has been found to be close to log-normal \citep{Stanek10,Fabjan11}. However, projection effects due to correlated structures and diffuse large-scale structure have been identified as an non-negligable source of scatter and bias in the mass scaling relation. The non-lognormal, positively skewed distribution of scatter in projected Compton $Y$ parameter in our cluster sample is in good agreement with the results of \citet{Hallman07} and \citet{Yang10}, who analyzed light cone/ cylindrical projections of the SZ effect, respectively. Based on an Edgeworth expansion of the mass - observable distribution, \citet{Shaw10} find that the higher order moments do not significantly impact the observed cluster mass function if the product of the scatter in the scaling relation, $\sigma_M$, and the slope of the mass function at the limiting mass of a survey is less than unity. Due to low scatter of the SZ scaling relation this criterion is met by all upcoming SZ experiments, suggesting that projection effects will be insignificant for cosmological constrains \citep[but see][for additional mitigation strategies]{Shaw08,Erickson11}.
\subsection{Influence of halo concentration}
\label{sec:cY}
The scatter in halo concentration at fixed cluster mass has been identified as an important source of scatter in X-ray temperature \citep{Yang09,Ameglio09} and SZ signal \citep{Shaw08,Yang10} of simulated clusters. Understanding the role of halo concentration on these observables is especially important for understanding selection biases and for the comparison to lensing derived cluster masses.\\
The right panel of Fig.~\ref{fig:yMz0} shows the correlation between scatter in halo concentration at fixed mass and scatter in $\lg Y_{200}$ at fixed mass for all clusters in sample B. We use the halo concentration measurements from \citet{Ameglio09} derived from fitting NFW-profiles to the integrated mass profile over the range $0.05< r/R_{\mr{vir}}< 1$, and model concentration $c(M_{200})$ with a power law in mass. The scatter is positively correlated with more concentrated clusters having higher SZ signals at fixed mass, with a correlation coefficient of 0.30 for the full sample B and 0.68 for the massive subsample B*. This result is in agreement with the positive correlation between scatter in concentration and spectroscopic-like temperature of these clusters reported in \citet{Ameglio09}. Similarly, \citet{Shaw08} find a positive correlation between scatter in concentration and integrated $Y$-parameter in halos from adiabatic SPH simulations and from N-body simulation in combination with semi-analytic gas models. On the other hand, \citet{Yang09,Yang10}  find a negative correlation between scatter in concentration\footnote{These authors use   $\lg(R_{200}/R_{500})$ as a proxy for concentration, which for an NFW profile is a monotonically decreasing function to halo concentration. We find correlation coefficients of -0.22 (-0.47) for the scatter in $\lg(R_{200}/R_{500})$ and $Y_{200}$ at fixed mass for sample B (B*), indicating that our result is robust with respect to the definition of halo concentration employed.} and scatter in temperature and integrated SZ signal. As discussed in \citet{Yang10}, the correlation between halo concentration and temperature at fixed mass depends on the assumed gas physics and the inclusion of radiative cooling, star formation and feedback may change the sign of the correlation.\\
On the observational side, \citet{Comerford10} find $\Delta T$ anticorrelated with $\Delta c$. However this analysis is based on a sample of 8 strong lensing clusters and the authors note that this result vanishes if a different measurement for the concentration of one cluster (MS 2137.3-2353) is used. As strong lensing selected cluster samples are strongly affected by projection effects and are biased towards higher halo concentrations and X-ray luminosities than average clusters \citep[e.g.][]{Meneghetti10, Meneghetti11}, larger, X-ray selected data sets like the CLASH survey \citep{Postman11} will be needed to observationally constrain the the correlation between scatter in temperature and halo concentration.\\

The scatter in halo concentration at fixed mass is linked to the formation epoch of a halo with more concentrated halos forming earlier \citep{NFW97}, albeit with large scatter \citep[e.g.][]{Neto07} which is likely due to enviromental effects \citep[see also][]{Gao07}. Hence the positive correlation between scatter in concentration and SZ signal suggests that clusters with $Y$ biased low formed more recently.
\begin{figure*}
\includegraphics[width = 0.95\textwidth, trim  = 0mm 0mm 0mm 0mm, clip = true]{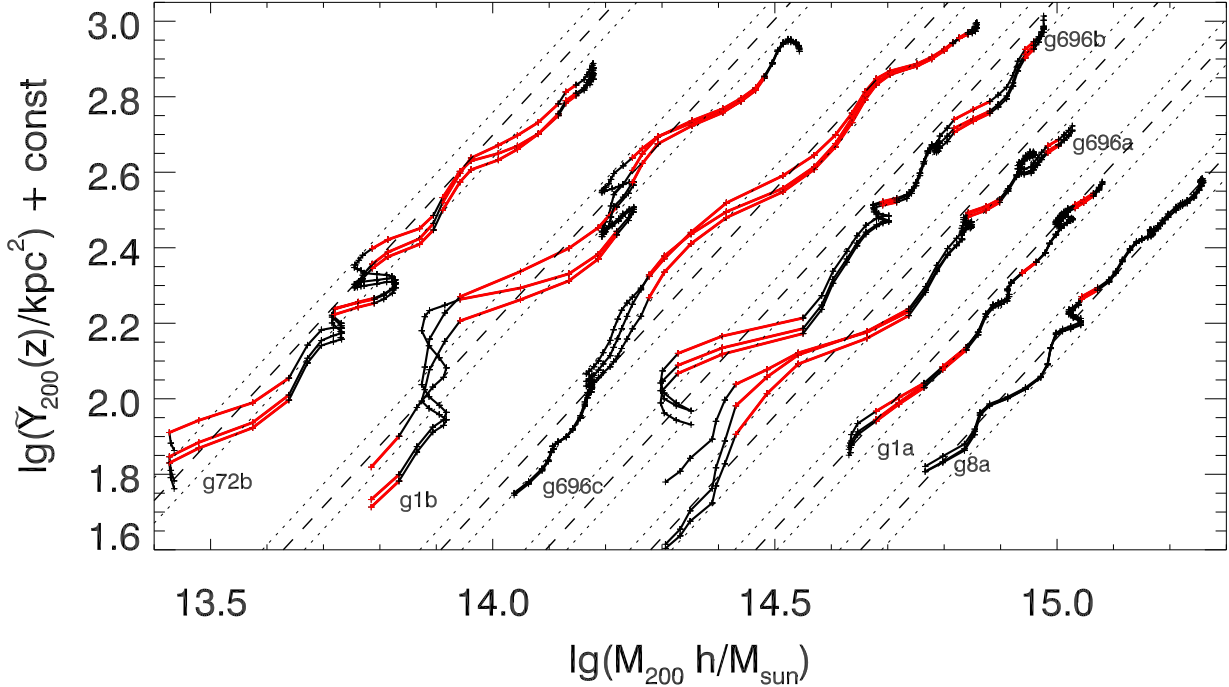}
\caption{Evolution of six massive clusters in mass and $\tilde Y_{200}$, the redshift evolution scaled $Y_{200}$. Offsets are added to show all clusters in one plot. We show three orthogonal projections for each cluster to illustrate the magnitude of projection effects. Phases identified as merging events are shown in red. The dashed and dotted lines show the best fit scaling relation for sample A and its $1\sigma$ error.}
\label{fig:evol}
\end{figure*}
\begin{figure*}
\includegraphics[width = 0.9\textwidth, trim  = 0mm 0mm 0mm 0mm, clip = true]{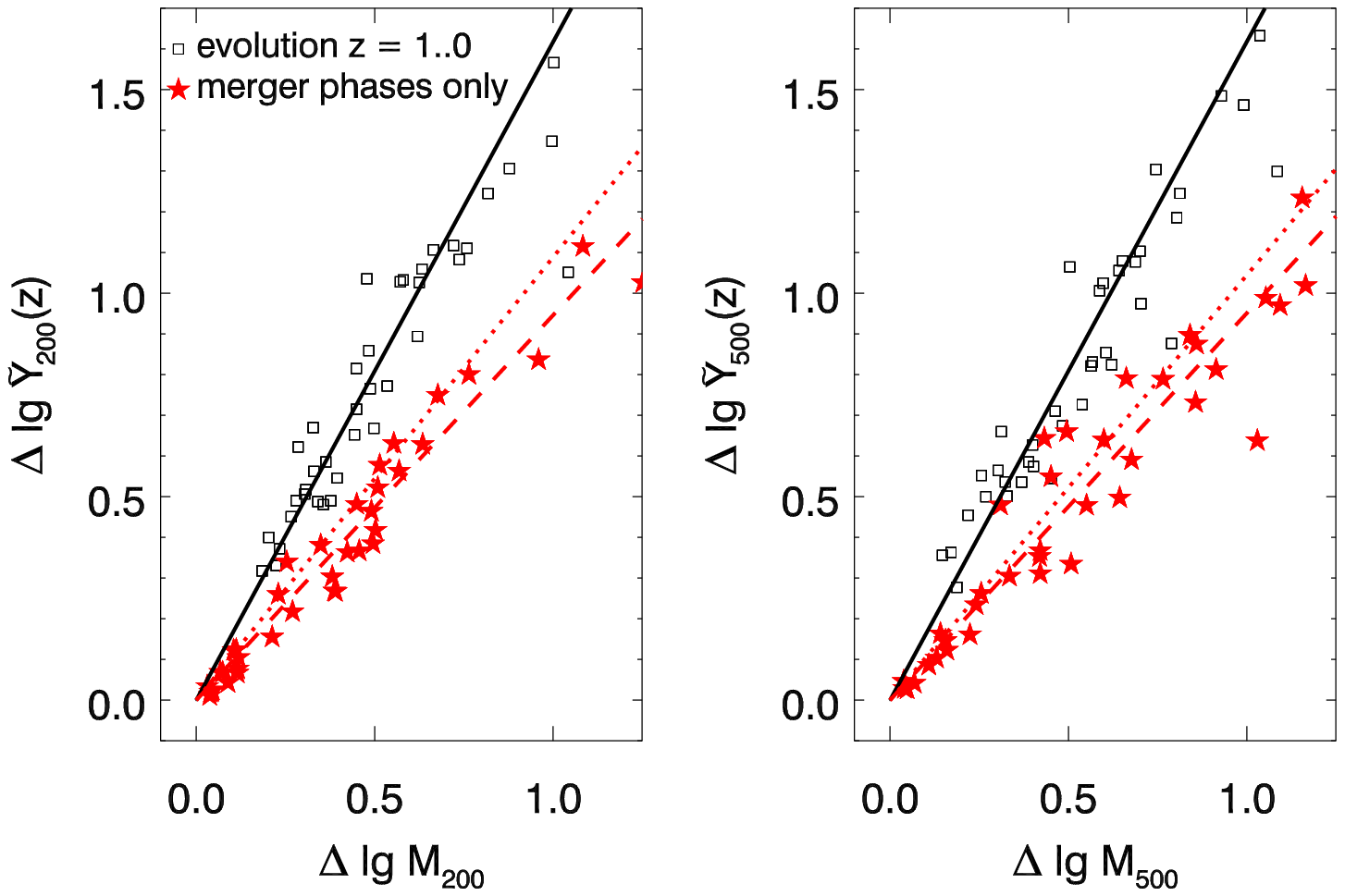}
\caption{Logarithmic mass growth and increase in SZ signal scaled for cosmological evolution for all clusters in sample A. The left panel shows the evolution within $\Delta = 200$, the right panel for $\Delta = 500$.The black open symbols show the overall evolution of individual clusters between $z=1$ and $z=0$, the black solid lines are the best linear fit with zero intercept to these points, yielding a slope of $1.62\pm0.19$ at $\Delta = 200$ ($1.62\pm0.29$ at $\Delta = 500$), consistent with the slope of the best fit scaling relation. Filled, red stars show the evolution of each cluster during merger phases, the dashed lines are the best linear fit with zero intercept to the evolution during mergers with slope  $0.94\pm0.15$ ($0.95\pm0.22$). The dotted lines show the best fit slope for the evolution during mergers when the merger criterion is relaxed to times when the fractional accretion rate per unit redshift is larger than the mean fractional accretion rate per unit redshift.}
\label{fig:evol2}
\end{figure*}
\section{Scatter induced by mergers}
\label{sec:merger}
We now turn to a detailed analysis of the evolution of merging clusters around the $M(Y)$ scaling relation fit to sample A. Figure~\ref{fig:evol} shows the trajectory of six massive clusters  around the best fit scaling relation in the $M_{200}$ -- $\tilde{Y}_{200}$ plane. Phases identified as mergers are shown in red. These examples suggest that the SZ signal lags behind the change in mass during extended merger events moving the merging clusters below the best fit scaling relation. This is similar to the findings of \citet{Rasia11} who analyzed the evolution of X-ray properties of two of these clusters (g8a and g1b) during mergers and find a time delay between mass increase and rise in temperature of order a few hundred mega years. We quantify the difference in evolution during mergers compared to the overall evolution of each cluster in the $M-Y$ plane in Fig.~\ref{fig:evol2}. The open symbols show the logarithmic increase in mass
\be
\Delta \lg M = \lg \left(\frac{M(z=0)}{ M(z=1)}\right)
\ee
and SZ signal scaled for redshift evolution
\be
\Delta \lg \tilde Y = \lg\left(\frac{ \tilde Y(z=0)}{\tilde Y(z=1)}\right)
\ee
As expected, the overall evolution from $z=1$ to $z=0$ as quantified by the slope of the best fit linear model with zero intercept is consistent with the slope of the best fit scaling relation.\\
The filled star symbols show the evolution of each cluster in the $M-\tilde Y$ plane during merger phases only (this corresponds to the sum of the red line segments for each cluster in Fig.~\ref{fig:evol}, treating the different projections separately). The dashed red lines indicate the best fit slope for the relation between increase in mass and redshift scaled $Y$ during mergers. This shows that the $Y$ signal scaled for redshift evolution increases more slowly during mergers than expected from the overall scaling relation. The dashed lines show the best fit slope for the relation between increase in mass and redshift scaled $Y$ during mergers when relaxing the merger criterion to include all times at which the fractional accretion rate is above its mean value. This illustrates that the suppression of $Y$ during mergers is robust with respect to the definition of merger event.\\

\begin{figure*}
\includegraphics[width = 0.9\textwidth, trim  = 0mm 0mm 0mm 0mm, clip = true]{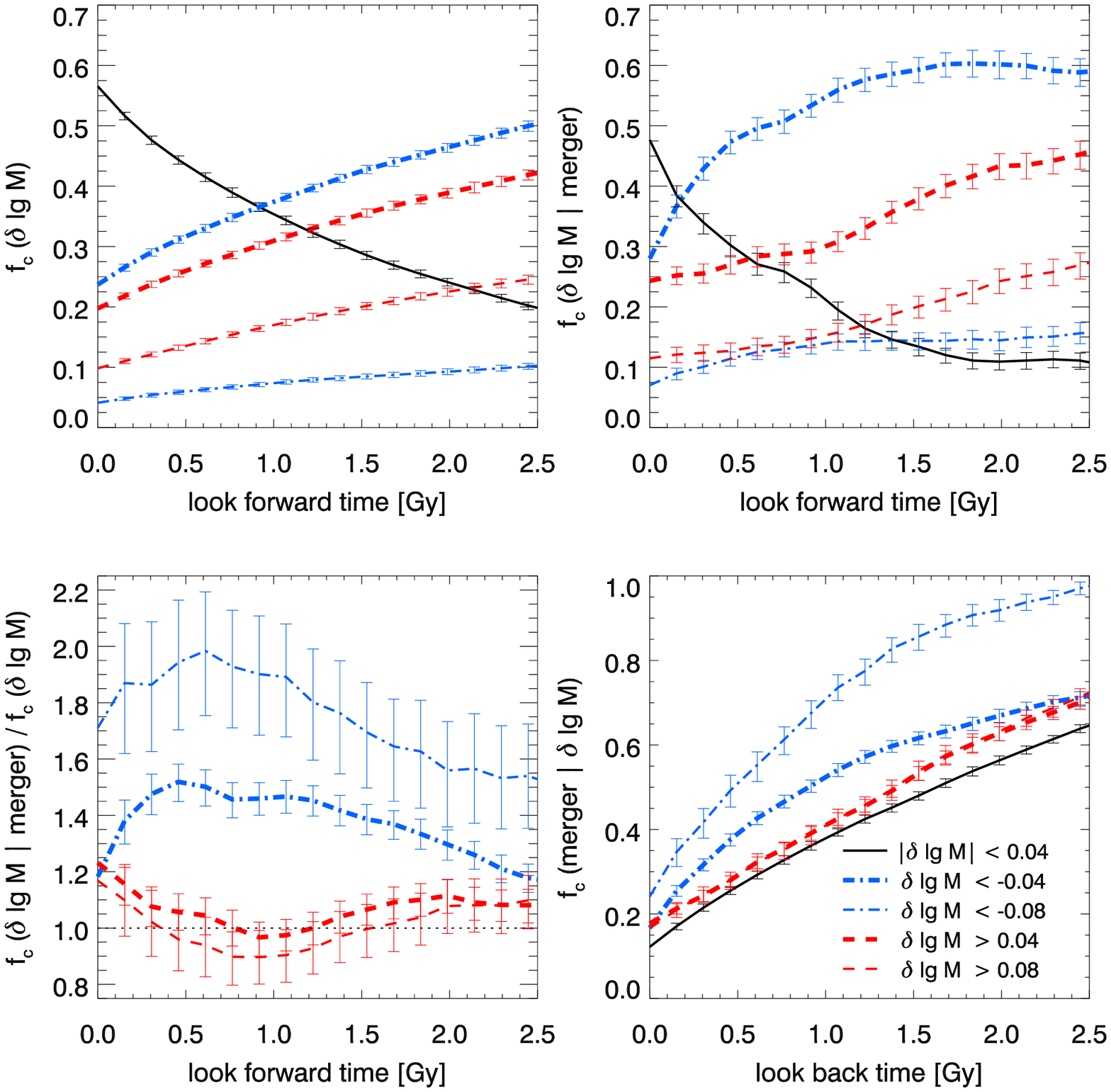}
\caption{Top left: Cumulative probability for a cluster to deviate from the scaling relation by $\delta \lg M$ as a function of time. Thick (thin) dash-dotted blue lines show the fraction of clusters deviating at least 0.04 (0.08) below the scaling relation, corresponding to a bias of 10\% (20\%) in the inferred mass. Thick (thin) dashed lines show the fraction of clusters deviating at least 0.04 (0.08) above the scaling relation. The black solid line show the fraction of cluster which deviate less that 10\% from the scaling relation within a given time. In all panels error bars indicate statistical errors estimated from 100 boot strap realizations.\newline
Top right: The same for merging clusters. Note that extended merging events are counted as multiple mergers, effectively giving more weight to major mergers.\newline
Bottom left: Ratio of the above panels, highlighting the enhanced probability for mergers to evolve below the scaling relation compared to an average cluster.\newline
Bottom right: Cumulative fraction of clusters which have undergone a merger as a function of look back time and their current deviation from the scaling relation.\newline
}
\label{fig:lookfw}
\end{figure*}
We further illustrate the connection between merging events and scatter in the $M_{200}(Y_{200})$ scaling relation in Fig.~\ref{fig:lookfw}. The top left panel shows how the clusters evolve around the scaling relation, giving the cumulative fraction of clusters evolving into outliers as a function of time, averaged over all clusters and all snapshots. Thick (thin) dashed-dotted or dashed lines show the fraction of clusters which evolve at least 10\% (20\%) below or above the scaling relation. For example, starting from one simulation snapshot, about 38\% of all clusters will move at least 10\% below the scaling relation within the next seven snapshots (corresponding to about one Gigayear), about 30\% deviate at least 10\% above the scaling relation during that time period and about 35\% stay within 10\% scatter from the scaling relation. The asymmetry between these pairs of lines is due to the non-lognormal distribution of scatter, the thick lines correspond to the 24\% and 80\% quantile, the thin lines correspond to the 4\% and 90\% quantile. The top right panel shows the same evolution around the scaling scaling for clusters undergoing a merger at $t=0$. Within a Gigayear after a merger, 55\% of all clusters will go through a phase where the inferred mass is biased low by at least 10\%, while for only 30\% of these cluster the inferred mass will be biased high by more than 10\% during this time. The bottom left panel shows the ratio of these two plots, and illustrates the asymmetric evolution of mergers below the scaling relation. The inferred mass of a recently merged cluster is about 50\% more likely to be biased low by at least 10\% and twice as likely to be biased low by at least 20\% compared to an average cluster.

The bottom right panel shows the cumulative fraction of clusters which have undergone a merger as a function of look back time given their current deviation from the scaling relation. This plot shows that 50\% (75\%) of all clusters with inferred masses biased low by at least 10\% (20\%) have undergone a merger within the last Gigayear.\\

In summary our analysis shows that the SZ signal changes more slowly than cluster mass during mergers. 
This indicates that for a cosmological distribution of merger orbits and mass ratios the delay between mass accretion and heating of the ICM by shocks and partial virialization are more important than merger boosts. 
Hence the inferred mass of recently merged clusters tends to be biased low and we find that a large fraction of negative outliers are associated with recent mergers.\\

Note that throughout this section we have analyzed deviations from a scaling relation determined from a fit to sample A. Since the merger histories of this environment selected sample are not necessarily representative of a volume limited sample the calibration of this relation may be biased. However, the results in this section and the correlation between scatter in halo concentration and SZ signal of the volume limited sample discussed in Sect.~\ref{sec:cY} suggest that this bias would increase the normalization $B$ and slope $\gamma$ at fixed $Y$. Hence such a calibration bias would downplay the asymmetric scatter induced by mergers that we reported in this section. This suggests that in a volume limited sample merging clusters may be less frequent, but their inferred masses could be more biased.
\begin{figure*}
 \centering
\includegraphics[width = 0.23\textwidth, trim  = 0mm 0mm 0mm 0mm, clip = true]{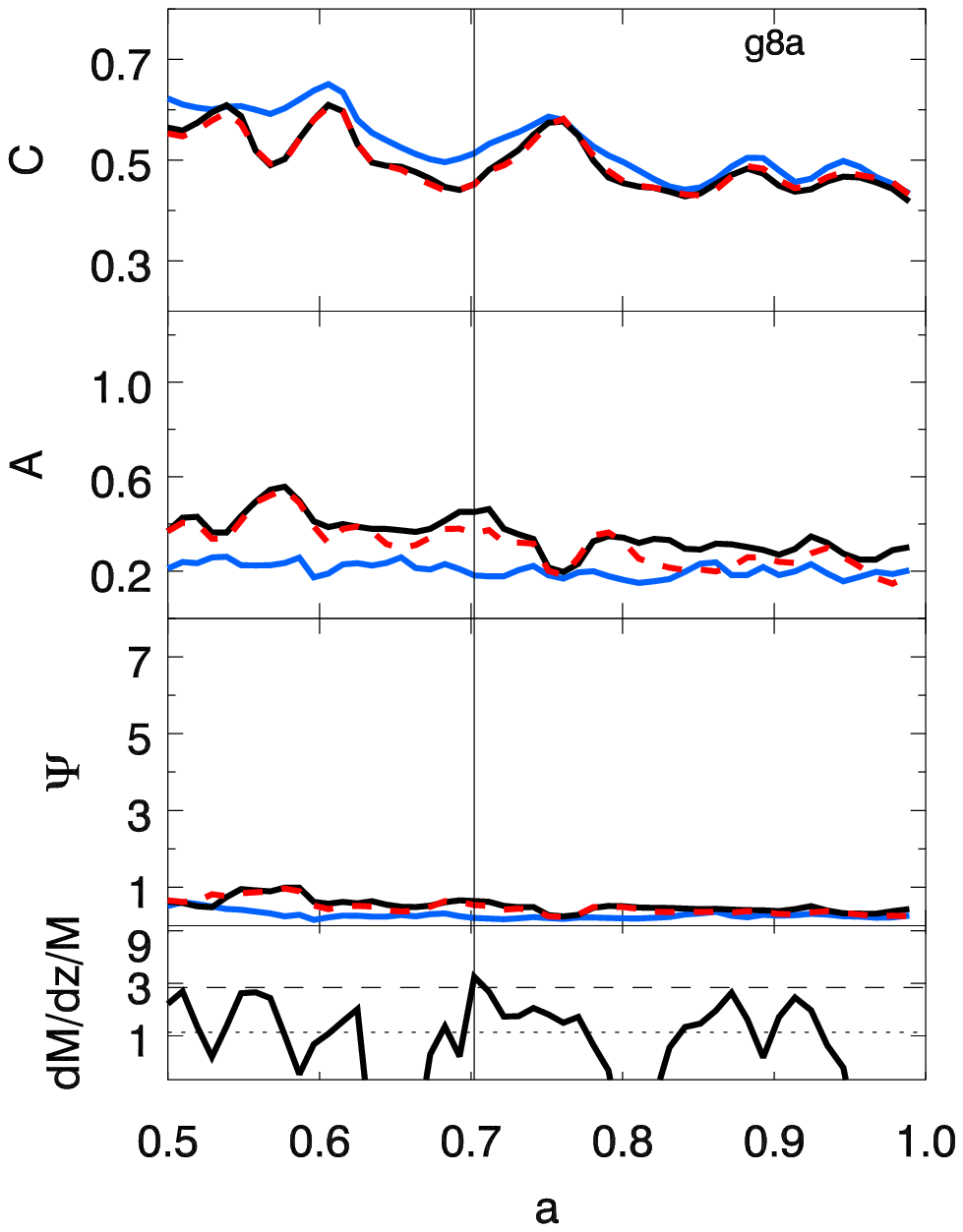}
\includegraphics[width = 0.23\textwidth, trim  = 0mm 0mm 0mm 0mm, clip = true]{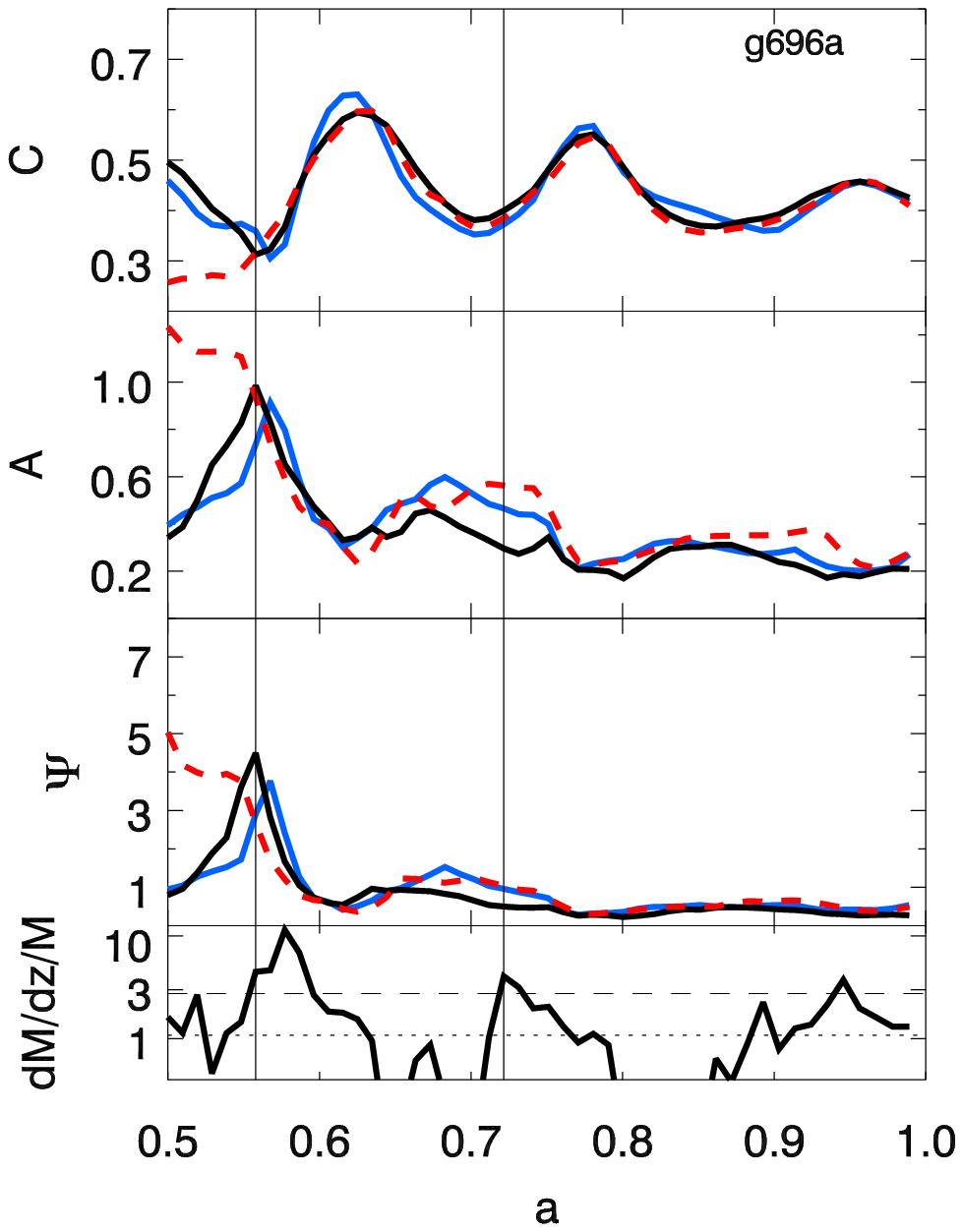}
\includegraphics[width = 0.23\textwidth, trim  = 0mm 0mm 0mm 0mm, clip = true]{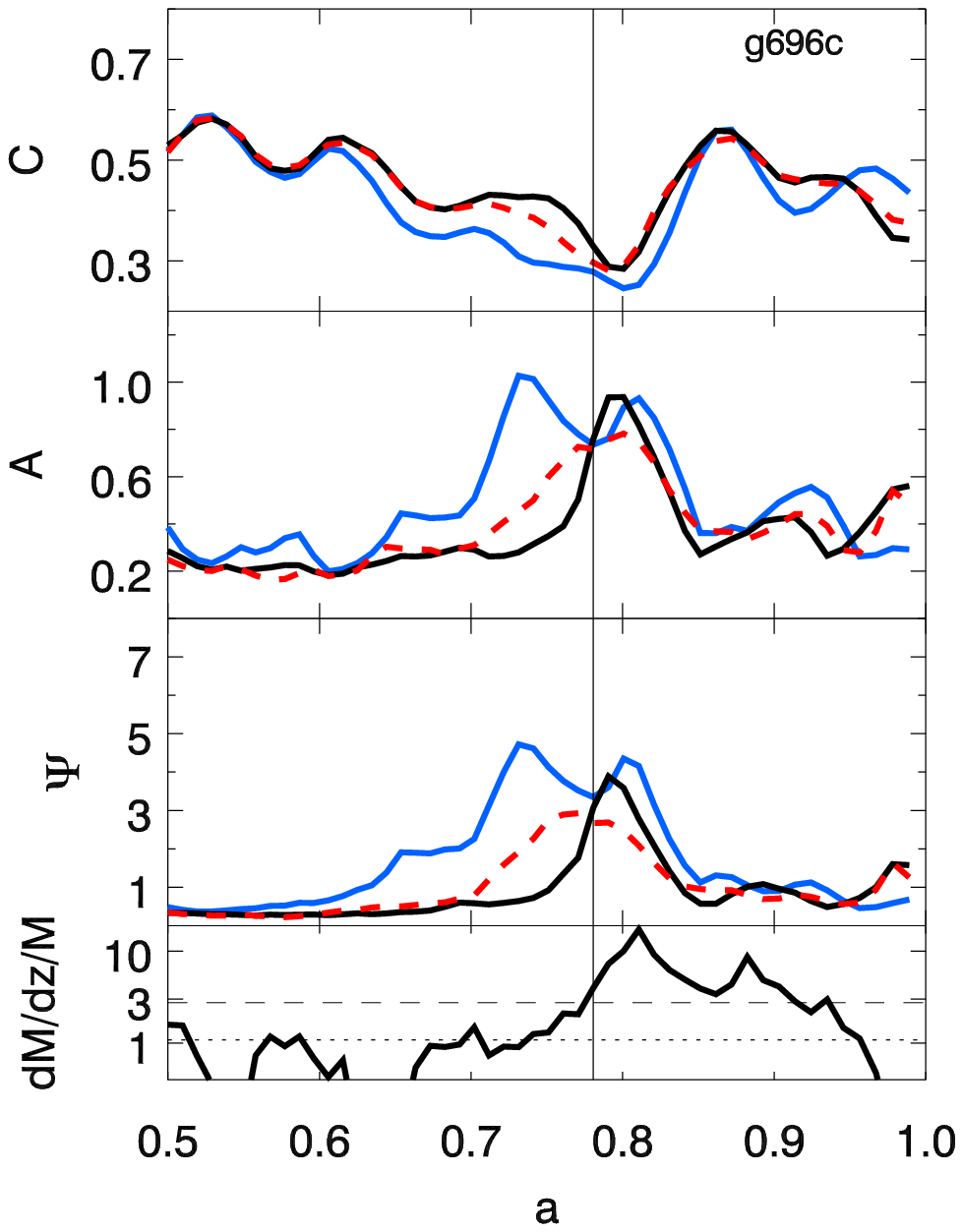}
\includegraphics[width = 0.23\textwidth, trim  = 0mm 0mm 0mm 0mm, clip = true]{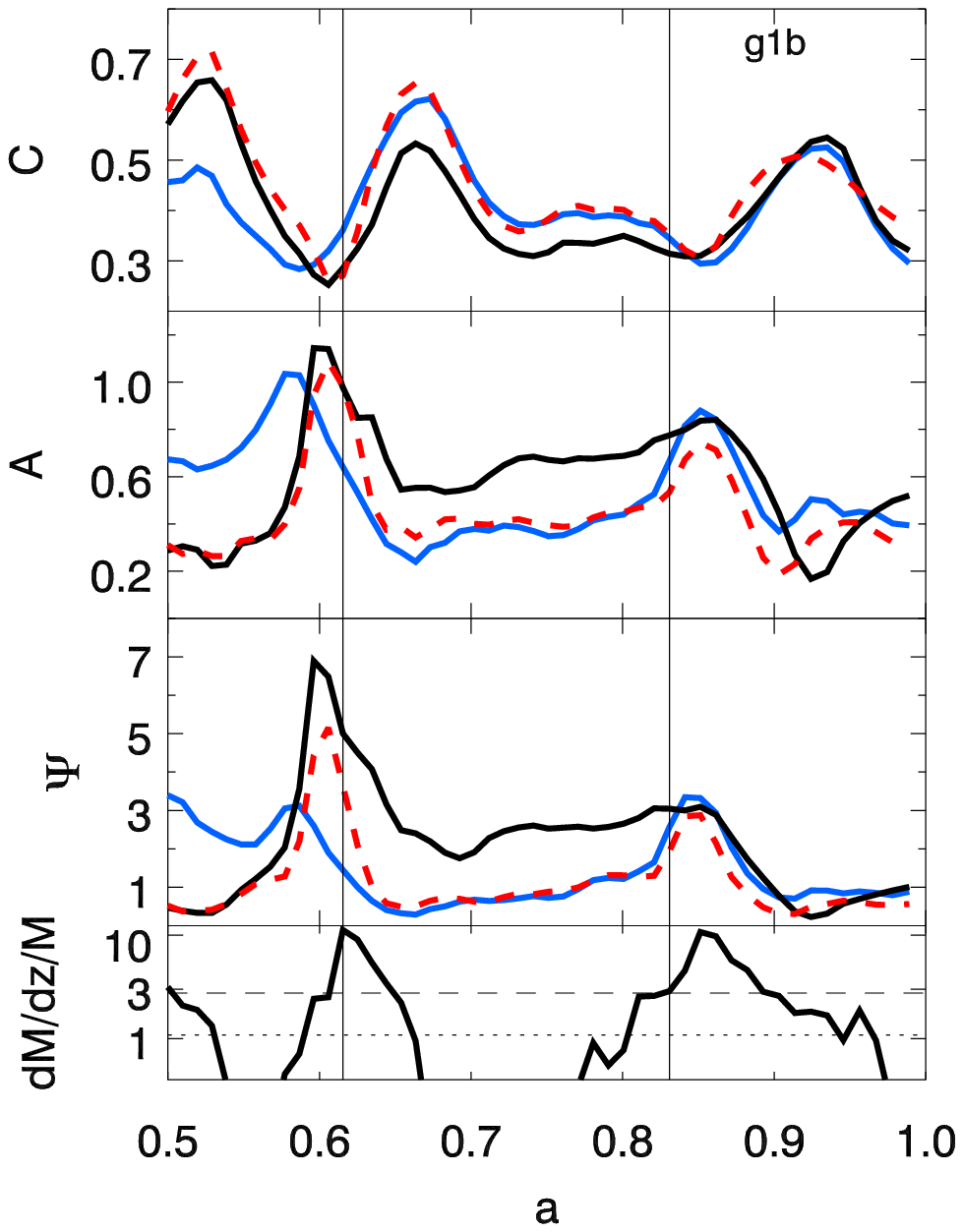}
\caption{Evolution of morphological parameters $G$, $A$, $\Psi$ for four massive clusters from sample A, different lines in each panel show the three orthogonal projections. The bottom panel shows the fractional accretion rate on a logarithmic scale, the dotted and dashed line indicate the mean accretion and the accretion rate threshold used to define mergers through out this analysis. Vertical lines mark the onset of mergers, i.e. the time when the fractional accretion rate first crosses the threshold used to define mergers. At the onset of a merger clusters appear less concentrated, more asymmetric and show more substructure.}
\label{fig:CASz}
\end{figure*}

\section{SZ Morphologies}
\label{sec:morph}
Since we found the dynamical state of clusters to be correlated with scatter in the $M(Y)$ scaling relation we now test if the morphological appearance of SZ maps can be used to identify clusters that deviate from the scaling relation. Quantitative measures of the X-ray surface brightness morphology are commonly used to identify disturbed clusters, observations \citep[e.g.][]{Boehringer10, Okabe10, Marrone11} and simulations \citep{Jeltema08, Ventimiglia08, Boehringer10} find the inferred masses of morphologically disturbed clusters to be biased low. \citet{Ventimiglia08} analyzed the morphology of clusters from the simulation of \citet{Borgani04}, which is our sample B, and find significant correlations between the centroid shift, axial ratio and power ratios of the X-ray surface brightness distribution of these clusters and scatter in the $T_{\mr X}(M)$ relation. \citet{Boehringer10} compared the morphology of these simulated clusters to observed morphologies in the REXCESS sample, and show that the simulated X-ray morphologies show a larger dynamic range and appear more disturbed during mergers. They trace this difference to the fact that cool cores are more pronounced in this simulation.\\

Here we test the effectiveness of a number of morphological parameters, which are typically used to measure X-ray morphology of clusters or optical morphology of galaxies, at quantifying substructure in projected $y$ maps. Within a circular aperture of radius $R_{200}$ we compute the following quantities:
\begin{itemize}
\item\emph{Asymmetry} $A$ measures substructures and differences from circular symmetry, it is defined as the normalized difference between an image $I$ and a copy $R$ of the image rotated by 180 degree, $A = \sum_{i}|I_{i}-R_{i}|/\sum_{i}I_{i}$, where sum runs over all pixels in the aperture, and the center of the aperture is chosen to minimize $A$ \citep{Conselice03}
\item \emph{Centroid shift} $w$ \citep{Mohr95} is another measure of the distribution of bright substructures based on the change of the centroid of different isophotal (iso-$y$) contours. Specifically, we follow the implementation of \citet{Ventimiglia08} and compute the variance of the centroid for 10 iso-$y$ contours spaced evenly in $\lg y$  between the maximum and minimum of $y$ within the aperture.
\item \emph{Concentration} $C$ We quantify the apparent concentration of the $y$ distribution by the fraction of integrated $Y$ contained within $0.3\times R_{200}$, $C= Y_{0.3R_{200}}/Y_{200}$
\item \emph{Ellipticity} $\epsilon= 1 -B/A$ is defined as the ratio of semi-major (A) and semi-minor axis (B) and is calculated directly from the second order moments of the $y$ distribution \citep{Hashimoto07}
\item \emph{Gini coefficient} $G$ measures the uniformness of pixel values regardless of their spatial distribution \citep{Lotz04}. It is based on the Lorentz curve, the rank--ordered cumulative distribution of pixel values. It is defined as
\be
G=\frac{1}{2\bar{y}n(n-1)}\sum_{i=1}^n\sum_{j=1}^n|y_i-y_j|\ ,
\ee
where $n$ is the number of pixels inside the aperture, $y_i$ the value of the $i$th pixel, and $\bar{y}$ is the mean pixel value. The Gini coefficient of a uniform distribution is zero, and it is one if one pixel contains all the signal. It increases with the fraction of $y$ in compact components.
\item \emph{Second order brightness moment} $M_{20}$ \citep{Lotz04}: The total second--order moment $M$ is the signal in each pixel $y_i$ weighted by the squared distance to the center of the galaxy cluster $(x_{1,c},x_{2,c})$, summed over all pixel inside the aperture:
\be
M=\sum_i^n M_i = \sum_i^n y_i \left((x_{1,i}-x_{1,c})^2+(x_{2,i}-x_{2,c})^2\right)\ .
\ee
Again, the center is determined by finding $(x_{1,c},x_{2,c})$ that minimizes $M$.
The second--order moment of the brightest regions measures the spatial distribution of bright sub clumps. $M_{20}$ is defined as the normalized second order moment of the brightest $20\%$ of the cluster's flux. $M_{20}$ is computed from the pixels rank ordered by $y$, 
\beq
M_{20}=\log\left(\frac{\sum_i M_i}{M}\right)&\rm{while}&\sum_i y_i < 0.2 Y_{200}\,.
\eeq
$M_{20}$ is similar to $C$, but it is more sensitive to the spatial distribution of luminous regions and is not based on any symmetry assumptions.
\item \emph{Multiplicity} $\Psi$ \citep{Law07} is another measure of the amount (multiplicity) of bright substructures. Using the observed $y$ distribution as a tracer of the cluster's projected mass, one can calculate a "potential energy" of the $y$ distribution 
\be
\Psi_{\rm{actual}} = \sum_{i=1}^n\sum_{j=1, j\neq i}^n\frac{y_i y_j}{r_{ij}}\ ,
\ee
where $r_{ij}$ is the distance between pixels $i$ and $j$. This value is normalized by the most compact possible re--arrangement of the pixel values, i.e. a circular configuration with pixel values decreasing with radius. The ``potential energy" of this most compact light distribution is
\be
\Psi_{\rm{compact}} = \sum_{i=1}^n\sum_{j=1, j\neq i}^n\frac{y_i y_j}{r^{\prime}_{ij}}\ ,
\ee
where $r^{\prime}_{ij}$ is the distance between pixels $i$ and $j$ in the most compact configuration.\\
The multiplicity coefficient is defined as
\be
\Psi = 100\times \log \left(\frac{\Psi_{\rm{compact}}}{\Psi_{\rm{actual}}}\right)\ .
\ee
It is similar to $A$ and $M_{20}$, but is has a larger dynamical range than $M_{20}$ and requires no center or symmetry assumption.

\item \emph{Power ratio} $P_n$ \citep{Buote95} correspond to a multipole expansion of the $y$ map inside an aperture centered on the $y$ centroid. We measure the power ratio $P_2/P_0$ which is related to the projected cluster ellipticity.
\end{itemize}
We measure morphology at a fixed physical resolution of 17.6 kpc/pixel and do not include any noise or observational effects.\\

Figure~\ref{fig:CASz} shows the morphology as measured by $C$, $A$, and $\Psi$ of four massive clusters from simulation A during their evolution since $a=0.5$. The evolution of these clusters around the $M(Y)$ scaling relation is shown in Fig.~\ref{fig:evol}. Vertical lines indicate the onset of mergers. Clusters g696a, g696c, and g1b illustrate the expected course of a merger: As a merging object enters the aperture within which morphologies are computed, the clusters appear less symmetric (higher $A$), less concentrated (lower $C$) and shows more substructure (higher $\Psi$). As the infalling clump sinks towards the cluster center and dissolves, the cluster appears less disturbed again. However, linking accretion history to morphology is complicated by extended merger phases (g696c, g1b at $a > 0.8$) with multiple infalling clumps. It is also apparent from these examples that fluctuation in morphology are not always linked to major accretion events (e.g. g8a, late time evolution of g696a).\\

\begin{figure*}
\centering
\includegraphics[width = \textwidth, trim  = 0mm 0mm 0mm 0mm, clip = true]{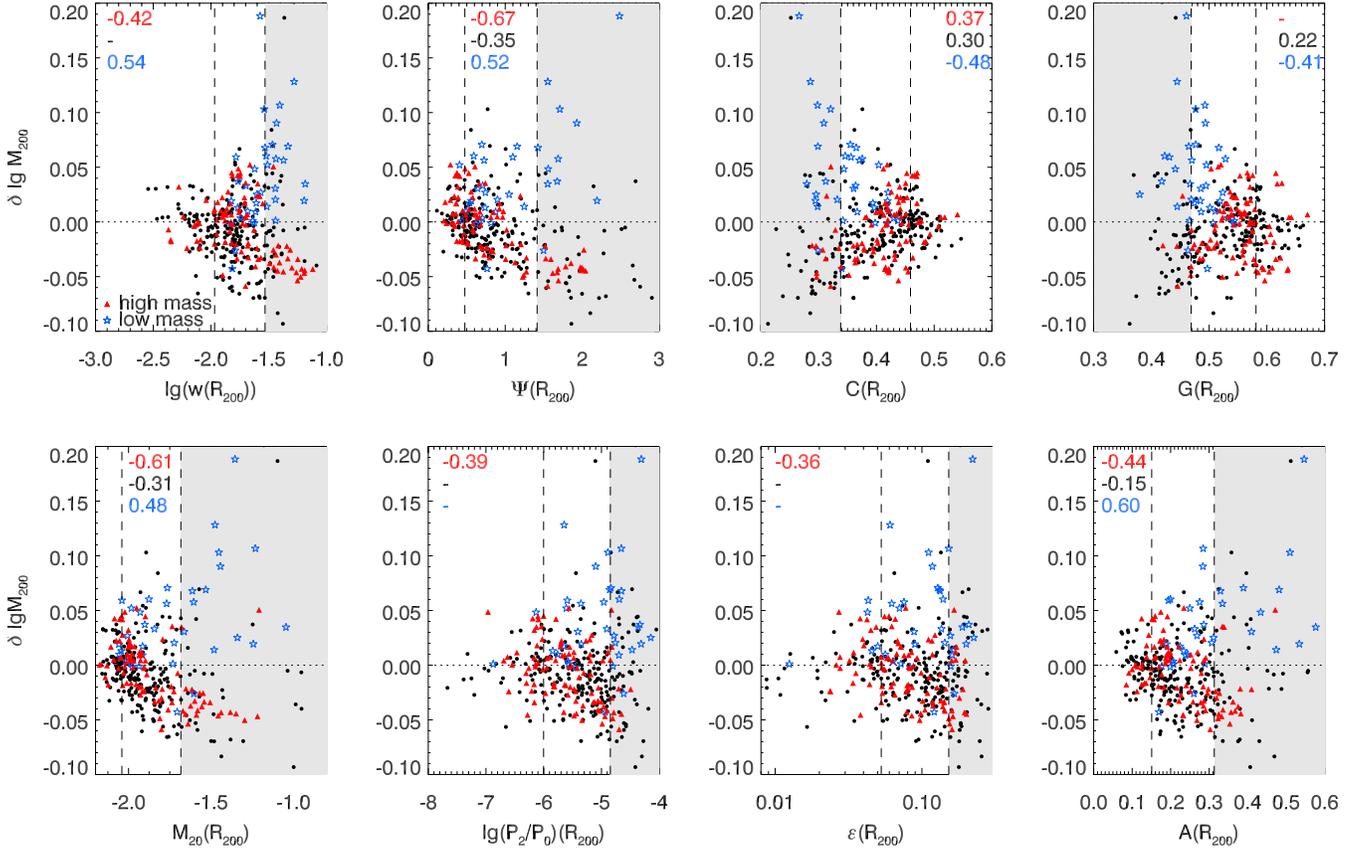}
\caption{Relation between scatter in the $M_{200}(Y_{200})$ relation $\delta \lg M_{200}$ and morphological parameters for all clusters from sample B measured within an aperture of size $R_{200}$. Open star symbols show clusters with $M<10^{14}M_{\odot}/h$, filled circles how clusters with $10^{14}M_{\odot}/h< M <2\times10^{14}M_{\odot}/h$, and filled triangles show massive clusters with $M> 2\times 10^{14} M_{\odot}/h$. Dashed vertical lines indicated the 25\% and 75\% quantiles of the morphology distribution. Shaded regions contain the 25\% of the data points which are classified as most disturbed by that morphological parameter. Numbers in the upper left or right corner give the Spearman rank correlation coefficient between the morphological parameter and scatter in the $M(Y)$ relation. From top to bottom these numbers are for mass samples $M>2\times10^{14}M_{\odot}/h$,$M>10^{14}M_{\odot}/h$,$M<10^{14}M_{\odot}/h$. If a correlation is not significant (significance level $>0.01),$ we do not list the correlation coefficient.}
\label{fig:CASz0}
\end{figure*}

\begin{figure*}
\centering
\includegraphics[width = \textwidth, trim  = 0mm 0mm 0mm 0mm, clip = true]{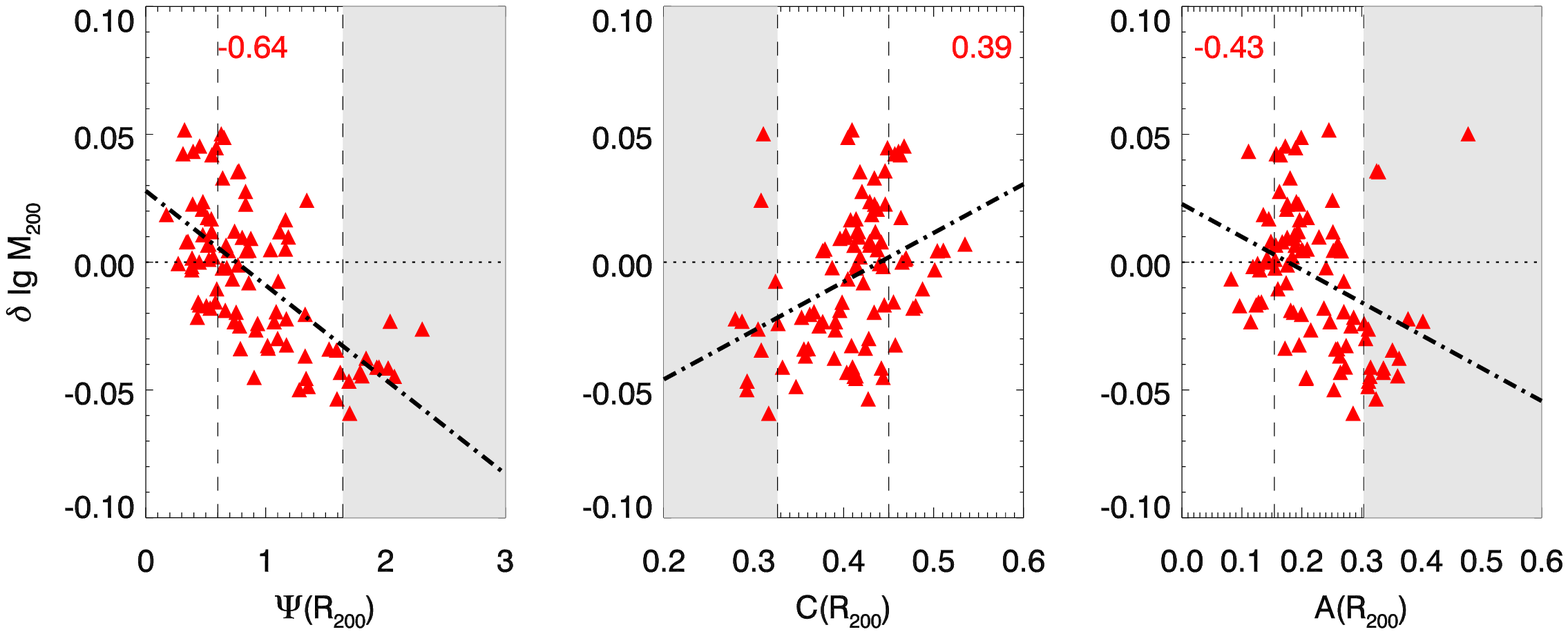}
\caption{Relation between scatter in the $M_{200}(Y_{200})$ relation $\delta \lg M_{200}$ and morphological parameters for clusters with $M>2\times10^{14}M_{\odot}/h$ from sample B, measured from SZ maps smoothed with a Gaussian beam with a FWHM of $150$ kpc and sampled at a pixel scale of $37.5$ kpc. Dashed vertical lines indicated the 25\% and 75\% quantiles of the morphology distribution. Shaded regions contain the 25\% of the data points which are classified as most disturbed by that morphological parameter. Numbers in the upper left or right corner give the Spearman rank correlation coefficient between the morphological parameter and scatter in the $M(Y)$ relation. Dashed-dotted lines show the best fit linear relation.}
\label{fig:CASz1}
\end{figure*}

For a more representative distribution of dynamical states and morphologies, we show the distribution of scatter in the $M(Y)$ relation and morphological parameters for all clusters in sample B in Fig.~\ref{fig:CASz0}. Shaded region contain the 25\% most disturbed/most elongated/least concentrated clusters. Overall, the inferred mass $M(Y)$ has larger scatter for clusters with disturbed morphologies, but it is nearly unbiased. Splitting the cluster sample by mass shows that morphologically disturbed clusters with low mass ($M_{200} < 10^{14}M_{\odot}/h$, open star symbols) tend to be biased towards larger inferred masses, while massive clusters ($M_{200} > 2\times 10^{14}M_{\odot}/h$, filled red triangles) with disturbed morphologies are preferentially biased low in inferred mass. We quantify this trend using the Spearman rank order correlation coefficient for different mass samples and show the correlation coefficients in Fig.~\ref{fig:CASz0}. If the significance level $s$ of a correlation between a morphology parameter and mass bias is low ($s> 0.01$), we do not list a correlation coefficient. We find a significant correlation between morphology and mass bias in all three mass bins ( $M>2\times10^{14}M_{\odot}/h$,$M>10^{14}M_{\odot}/h$,$M<10^{14}M_{\odot}/h$) for the multiplicity, concentration,$M_{20}$ and asymmetry parameter. These different morphology parameters consistently show that the correlation between disturbed morphology and negative mass bias increases with mass threshold, and the correlation coefficient changes sign for the low mass clusters. For centroid shifts and the Gini coefficient, we only find significant correlations with scatter in the $M(Y)$ relation in two mass bins, which follow the same pattern as just described. Power ratio $P_2/P_0$ and ellipticity are correlated with mass bias only for the most massive clusters, such that less circular clusters tend to be biased low in mass.\\
This segregation in mass, which is consistent among all morphological parameters, suggests that a large fraction of morphologically disturbed clusters which are biased high in inferred mass is caused by projection effects. The more massive clusters, which are less affected by projection effects, show correlations with disturbed morphology corresponding to a negative bias in inferred mass as expected from X-ray results. We expect cool cores to have a smaller influence on the SZ morphology than is found in X-ray, as the SZ signal is linear in density and less sensitive to physics in the cluster core. Projection effects due to uncorrelated large scale structure along the line of sight are on average more diffuse than the projection effects from nearby structure that is included in our analysis. Hence we do not expect the morphology of massive clusters to become dominated by projection effects for line of sight projections which include all intervening structure.\\
As a first step towards towards including resolution effects, we convolve all projected $y$ maps with a circular Gaussian beam with full width at half maximum (FWHM) of 150 kpc, and sample the maps at a resolution of four pixels per FWHM. For a telescope with an 1 arc minute beam, this physical resolution is reached for a source at $z\sim0.15$; for an experiment with beam width of about 20 arc seconds, this corresponds to $z\sim 0.8$. Figure~\ref{fig:CASz1} shows the correlation between mass bias and cluster morphology as measured from these blurred maps for all massive clusters with $M>2\times10^{14}M_{\odot}/h$ from sample B. For this choice of beam and pixel scale, cluster morphology and bias in inferred mass are well correlated and resolution effects are small. However, since this analysis is based on noise- and background-free $y$ maps and a simplistic map making procedure, more realistic simulations are required to assess whether SZ based morphology can in practice be used as a proxy for the dynamical state of a cluster.
\section{Summary and Discussion}
\label{sec:summary}
Using projected Compton $y$ maps of galaxy clusters extracted from cosmological hydrodynamical simulations, we analyze the clusters' thermal SZ signal and its scaling relation with cluster mass. We study the detailed time evolution of a sample of 39 clusters around the scaling relation using simulations with outputs closely spaced in time. Compared to previous studies, which focused either on the evolution of isolated, idealized mergers or on large samples of clusters at widely spaced redshifts, this sample enables us to isolate the effect of merging events for a cosmologically representative distribution of merger orbits, mass ratios, and impact parameters. Our main results can be summarized as follows:
\begin{enumerate}
\item The best fit scaling relations to the integrated $Y_{200}$ signal of these clusters are close to self-similar predictions and agree well with other simulations that include comparable gas physics.
\item The scatter around these scaling relations is small (of order 10\% scatter in mass at fixed $Y_{200}$) and it is overall well correlated with the scatter in halo concentration, such that more concentrated halos have larger $Y$ signal at fixed mass.
\item The scatter in the scaling relation deviates from a log normal distribution and is skewed towards clusters with $Y$ signals larger than expected from their mass. We find projection effects due to nearby structures to be an important source of this upward scatter. However, due to the small magnitude of the scatter in the mass scaling projection effects are not expected to be a significant contamination for cosmological constraints from SZ cluster surveys. 
\item Merging clusters fall below the scaling relation, such that their inferred masses are biased low. More quantitatively, we find that within a Gyr following a merger, clusters are twice as likely as the average cluster to undergo a phase during which their inferred mass is biased low by more than 10\%. 
\item We identify merging events to be a major source of downward scatter in the scaling relation: a large fraction of clusters whose inferred masses are biased low recently underwent a merger (c.f. Fig.~\ref{fig:lookfw}).
\item For massive clusters, we find the morphology of SZ maps to be well correlated with deviations from the scaling relation. While the robustness of this result with respect to noise and imaging artifacts requires further analysis, it suggests that SZ morphology may be useful to reduce the scatter of mass estimates, and to infer merger rates of massive halos and hence test theories of halo formation.
\end{enumerate}
Our analysis of the time evolution of merging events is in agreement with the conclusions drawn from earlier studies comparing morphologically disturbed and undisturbed clusters in cosmological simulations at fixed redshifts \citep[e.g.][]{Mathiesen01, Kravtsov06,Nagai06,Jeltema08, Ventimiglia08}. Specifically, it supports the hypothesis that for a cosmological distribution of merger parameters partial virialization and non-thermal pressure support due to mergers are more important than merger boosts found in simulations of direct collisions between mergers. For simulated clusters the intrinsic scatter in the scaling relation and the mass segregation between morphologically relaxed and disturbed clusters are significantly smaller than recent observational results based on SZ measurements, X-ray morphology and weak lensing inferred masses \citep{Marrone11}. However, as these authors note, the observed scatter is in agreement with the scatter expected in weak lensing mass measurements \citep{Becker10}. Similarly, the mass segregation is enhanced by the sensitivity of weak lensing mass estimates to cluster triaxiality, and these observational constraints on the intrinsic scatter and bias in SZ mass estimates are limited by the accuracy of weak lensing mass reconstruction.\\

Further complications arise when inferring cluster masses from SZ observations as most $Y$ measurements are derived from fitting parametric profiles \citep[e.g.][]{Nagai07p,Arnaud10} to the data which assume radial symmetry \citep[but see][for alternate methods and discussions]{Plagge10,Marrone11,Sayers11}.The distorted geometry of merging clusters may introduce additional scatter to mass estimates derived from profile fits, but an experiment specific analysis of such effects is beyond the scope of this work. \\

An additional limitation of our analysis is the range of non-gravitational physics included in the simulations. While recent studies show the impact of AGN-feedback on overall cluster profiles and scaling relations \citep{Sijacki07,Puchwein08,Battaglia10, Fabjan11}, this mainly affects the cluster center. Consequently, we do not expect AGN feedback to significantly alter the slow virialization of newly accreted material at larger radii, which we found to be the main source of scatter during merging events.
In the cluster outskirts, electrons and ions are not in thermal equilibrium. \citet{Rudd09} and \citet{Wong09} show that detailed treatment of the multi-temperature structure of the intracluster medium leads to a significant suppression of electron temperature and SZ signal. Based on a sample of three simulated cluster, \citet{Rudd09} find this effect to be especially pronounced in clusters undergoing major mergers. Under specific conditions, this effect may cause a bias of up to $5\%$ in integrated $Y$, corresponding to an additional negative bias of about $ 3\%$ in the inferred mass of merging clusters.\\

Overall, we find that merger events cause a temporary negative bias in inferred cluster mass of order $10\%-15\%$. Due to the increased fraction of recently merged objects at higher redshift, we conclude that this merger bias should be accounted for when modeling the redshift evolution in the scatter of scaling relations.
\section*{Acknowledgements}
We thank Silvia Ameglio for providing SZ maps and halo concentration measurements for cluster sample B, and Nick Battaglia, Chris Hirata, and James Taylor for useful discussions.
The JobRunner web application was constructed by Laurent Bourges and Gerard Lemson as part of the activities of the German Astrophysical Virtual Observatory. EK is supported by the US National Science Foundation (AST-0807337), the US Department of Energy (DE-FG03-02-ER40701), and the David and Lucile
Packard Foundation. EP acknowledges support from NASA
grant NNX07AH59G and JPL Planck subcontract 1290790. EP and EK were supported by NSF grant AST-0649899 during the early stages of this work. K.D. acknowledges the support by the DFG Priority Programme 1177 and
additional support by the DFG Cluster of Excellence "Origin and Structure of the Universe".
\bibliographystyle{aa}


\begin{thebibliography}{}

\bibitem[\protect\citeauthoryear{{Allen}, {Evrard} \& {Mantz}}{{Allen}
  et~al.}{2011}]{Allen11}
{Allen} S.~W.,  {Evrard} A.~E.,    {Mantz} A.~B.,  2011, ArXiv e-prints

\bibitem[\protect\citeauthoryear{{Ameglio}, {Borgani}, {Pierpaoli} \&
  {Dolag}}{{Ameglio} et~al.}{2007}]{Ameglio07}
{Ameglio} S.,  {Borgani} S.,  {Pierpaoli} E.,    {Dolag} K.,  2007, \mnras,
  382, 397

\bibitem[\protect\citeauthoryear{{Ameglio}, {Borgani}, {Pierpaoli} \& et
  al.}{{Ameglio} et~al.}{2009}]{Ameglio09}
{Ameglio} S.,  {Borgani} S.,  {Pierpaoli} E.,    et al. 2009, \mnras, 394, 479

\bibitem[\protect\citeauthoryear{{Andersson}, {Benson}, {Ade} \& et
  al.}{{Andersson} et~al.}{2010}]{Andersson10}
{Andersson} K.,  {Benson} B.~A.,  {Ade} P.~A.~R.,    et al. 2010, ArXiv
  e-prints

\bibitem[\protect\citeauthoryear{{Angrick} \& {Bartelmann}}{{Angrick} \&
  {Bartelmann}}{2011}]{Angrick11}
{Angrick} C.,  {Bartelmann} M.,  2011, ArXiv e-prints

\bibitem[\protect\citeauthoryear{{Arnaud}, {Pratt}, {Piffaretti} \& et
  al.}{{Arnaud} et~al.}{2010}]{Arnaud10}
{Arnaud} M.,  {Pratt} G.~W.,  {Piffaretti} R.,    et al. 2010, \aap, 517, A92+

\bibitem[\protect\citeauthoryear{{Battaglia}, {Bond}, {Pfrommer},  \& et
  al.}{{Battaglia} et~al.}{2010}]{Battaglia10}
{Battaglia} N.,  {Bond} J.~R.,  {Pfrommer} C.,     et al. 2010, \apj, 725, 91

\bibitem[\protect\citeauthoryear{{Becker} \& {Kravtsov}}{{Becker} \&
  {Kravtsov}}{2010}]{Becker10}
{Becker} M.~R.,  {Kravtsov} A.~V.,  2010, ArXiv e-prints

\bibitem[\protect\citeauthoryear{{B{\"o}hringer}, {Pratt}, {Arnaud} \& et
  al.}{{B{\"o}hringer} et~al.}{2010}]{Boehringer10}
{B{\"o}hringer} H.,  {Pratt} G.~W.,  {Arnaud} M.,    et al. 2010, \aap, 514,
  A32+

\bibitem[\protect\citeauthoryear{{Borgani}, {Murante}, {Springel},  \& et
  al.}{{Borgani} et~al.}{2004}]{Borgani04}
{Borgani} S.,  {Murante} G.,  {Springel} V.,     et al. 2004, \mnras, 348, 1078

\bibitem[\protect\citeauthoryear{{Buote} \& {Tsai}}{{Buote} \&
  {Tsai}}{1995}]{Buote95}
{Buote} D.~A.,  {Tsai} J.~C.,  1995, \apj, 452, 522

\bibitem[\protect\citeauthoryear{{Comerford}, {Moustakas} \&
  {Natarajan}}{{Comerford} et~al.}{2010}]{Comerford10}
{Comerford} J.~M.,  {Moustakas} L.~A.,    {Natarajan} P.,  2010, \apj, 715, 162

\bibitem[\protect\citeauthoryear{{Conselice}}{{Conselice}}{2003}]{Conselice03}
{Conselice} C.~J.,  2003, \apjs, 147, 1

\bibitem[\protect\citeauthoryear{{Dolag}, {Borgani}, {Murante} \&
  {Springel}}{{Dolag} et~al.}{2009}]{Dolag09}
{Dolag} K.,  {Borgani} S.,  {Murante} G.,    {Springel} V.,  2009, \mnras, 399,
  497

\bibitem[\protect\citeauthoryear{{Dolag}, {Hansen}, {Roncarelli} \&
  {Moscardini}}{{Dolag} et~al.}{2005}]{Dolag05}
{Dolag} K.,  {Hansen} F.~K.,  {Roncarelli} M.,    {Moscardini} L.,  2005,
  \mnras, 363, 29

\bibitem[\protect\citeauthoryear{{Dolag}, {Meneghetti}, {Moscardini} \& et
  al.}{{Dolag} et~al.}{2006}]{Dolag06}
{Dolag} K.,  {Meneghetti} M.,  {Moscardini} L.,    et al. 2006, \mnras, 370,
  656

\bibitem[\protect\citeauthoryear{{Erickson}, {Cunha} \& {Evrard}}{{Erickson}
  et~al.}{2011}]{Erickson11}
{Erickson} B.~M.~S.,  {Cunha} C.~E.,    {Evrard} A.~E.,  2011, ArXiv e-prints

\bibitem[\protect\citeauthoryear{{Fabjan}, {Borgani}, {Rasia}, {Bonafede} \&
  {Dolag}}{{Fabjan} et~al.}{2011}]{Fabjan11}
{Fabjan} D.,  {Borgani} S.,  {Rasia} E.,  {Bonafede} A.,    {Dolag} K.,  2011,
  ArXiv e-prints

\bibitem[\protect\citeauthoryear{{Fakhouri} \& {Ma}}{{Fakhouri} \&
  {Ma}}{2008}]{Fakhouri08}
{Fakhouri} O.,  {Ma} C.,  2008, \mnras, 386, 577

\bibitem[\protect\citeauthoryear{{Fakhouri} \& {Ma}}{{Fakhouri} \&
  {Ma}}{2009}]{Fakhouri09}
{Fakhouri} O.,  {Ma} C.,  2009, \mnras, 394, 1825

\bibitem[\protect\citeauthoryear{{Gao}, {Springel} \& {White}}{{Gao}
  et~al.}{2005}]{Gao05}
{Gao} L.,  {Springel} V.,    {White} S.~D.~M.,  2005, \mnras, 363, L66

\bibitem[\protect\citeauthoryear{{Gao} \& {White}}{{Gao} \&
  {White}}{2007}]{Gao07}
{Gao} L.,  {White} S.~D.~M.,  2007, \mnras, 377, L5

\bibitem[\protect\citeauthoryear{{Hallman},{O'Shea}, {Burns} \& et al.}{{Hallman} et~al.}{2007}]{Hallman07}
{Hallman} E.~J.,  {O'Shea} B.~W.,{Burns} J.~O., et al.  2007, \apj, 671, 27

\bibitem[\protect\citeauthoryear{{Hashimoto}, {B{\"o}hringer}, {Henry},
  {Hasinger} \& {Szokoly}}{{Hashimoto} et~al.}{2007}]{Hashimoto07}
{Hashimoto} Y.,  {B{\"o}hringer} H.,  {Henry} J.~P.,  {Hasinger} G.,
  {Szokoly} G.,  2007, \aap, 467, 485

\bibitem[\protect\citeauthoryear{{Jeltema}, {Hallman}, {Burns} \&
  {Motl}}{{Jeltema} et~al.}{2008}]{Jeltema08}
{Jeltema} T.~E.,  {Hallman} E.~J.,  {Burns} J.~O.,    {Motl} P.~M.,  2008,
  \apj, 681, 167

\bibitem[\protect\citeauthoryear{{Kravtsov}, {Vikhlinin} \& {Nagai}}{{Kravtsov}
  et~al.}{2006}]{Kravtsov06}
{Kravtsov} A.~V.,  {Vikhlinin} A.,    {Nagai} D.,  2006, \apj, 650, 128

\bibitem[\protect\citeauthoryear{{Lau}, {Kravtsov} \& {Nagai}}{{Lau}
  et~al.}{2009}]{Lau09}
{Lau} E.~T.,  {Kravtsov} A.~V.,    {Nagai} D.,  2009, \apj, 705, 1129

\bibitem[\protect\citeauthoryear{{Law}, {Steidel}, {Erb} \& et al.}{{Law}
  et~al.}{2007}]{Law07}
{Law} D.~R.,  {Steidel} C.~C.,  {Erb} D.~K.,    et al. 2007, \apj, 656, 1

\bibitem[\protect\citeauthoryear{{Lima} \& {Hu}}{{Lima} \& {Hu}}{2005}]{Lima05}
{Lima} M.,  {Hu} W.,  2005, \prd, 72, 043006

\bibitem[\protect\citeauthoryear{{Lotz}, {Primack} \& {Madau}}{{Lotz}
  et~al.}{2004}]{Lotz04}
{Lotz} J.~M.,  {Primack} J.,    {Madau} P.,  2004, \aj, 128, 163

\bibitem[\protect\citeauthoryear{{Majumdar} \& {Mohr}}{{Majumdar} \&
  {Mohr}}{2004}]{Majumdar04}
{Majumdar} S.,  {Mohr} J.~J.,  2004, \apj, 613, 41

\bibitem[\protect\citeauthoryear{{Marriage}, {Acquaviva}, {Ade} \& et
  al.}{{Marriage} et~al.}{2010}]{Marriage10}
{Marriage} T.~A.,  {Acquaviva} V.,  {Ade} P.~A.~R., et al. 2010, ArXiv
  e-prints

\bibitem[\protect\citeauthoryear{{Marrone}, {Smith}, {Okabe} \& et
  al.}{{Marrone} et~al.}{2011}]{Marrone11}
{Marrone} D.~P.,  {Smith} G.~P.,  {Okabe} N.,    et al. 2011, ArXiv
  e-prints

\bibitem[\protect\citeauthoryear{{Mathiesen} \& {Evrard}}{{Mathiesen} \&
  {Evrard}}{2001}]{Mathiesen01}
{Mathiesen} B.~F.,  {Evrard} A.~E.,  2001, \apj, 546, 100

\bibitem[\protect\citeauthoryear{{McBride}, {Fakhouri} \& {Ma}}{{McBride}
  et~al.}{2009}]{McBride09}
{McBride} J.,  {Fakhouri} O.,    {Ma} C.,  2009, \mnras, 398, 1858

\bibitem[\protect\citeauthoryear{{Meneghetti}, {Fedeli}, {Pace},
  {Gottl{\"o}ber} \& {Yepes}}{{Meneghetti} et~al.}{2010}]{Meneghetti10}
{Meneghetti} M.,  {Fedeli} C.,  {Pace} F.,  {Gottl{\"o}ber} S.,    {Yepes} G.,
  2010, \aap, 519, A90+

\bibitem[\protect\citeauthoryear{{Meneghetti}, {Fedeli}, {Zitrin},
  {Bartelmann}, {Broadhurst}, {Gottl{\"o}ber}, {Moscardini} \&
  {Yepes}}{{Meneghetti} et~al.}{2011}]{Meneghetti11}
{Meneghetti} M.,  {Fedeli} C.,  {Zitrin} A.,  {Bartelmann} M.,  {Broadhurst}
  T.,  {Gottl{\"o}ber} S.,  {Moscardini} L.,    {Yepes} G.,  2011, \aap, 530,
  A17+

\bibitem[\protect\citeauthoryear{{Mohr}, {Evrard}, {Fabricant} \&
  {Geller}}{{Mohr} et~al.}{1995}]{Mohr95}
{Mohr} J.~J.,  {Evrard} A.~E.,  {Fabricant} D.~G.,    {Geller} M.~J.,  1995,
  \apj, 447, 8

\bibitem[\protect\citeauthoryear{{Nagai}}{{Nagai}}{2006}]{Nagai06}
{Nagai} D.,  2006, \apj, 650, 538

\bibitem[\protect\citeauthoryear{{Nagai}, {Kravtsov} \& {Vikhlinin}}{{Nagai}
  et~al.}{2007a}]{Nagai07p}
{Nagai} D.,  {Kravtsov} A.~V.,    {Vikhlinin} A.,  2007a, \apj, 668, 1

\bibitem[\protect\citeauthoryear{{Nagai}, {Vikhlinin} \& {Kravtsov}}{{Nagai}
  et~al.}{2007b}]{Nagai07}
{Nagai} D.,  {Vikhlinin} A.,    {Kravtsov} A.~V.,  2007b, \apj, 655, 98

\bibitem[\protect\citeauthoryear{{Navarro}, {Frenk} \& {White}}{{Navarro}
  et~al.}{1997}]{NFW97}
{Navarro} J.~F.,  {Frenk} C.~S.,    {White} S.~D.~M.,  1997, \apj, 490, 493

\bibitem[\protect\citeauthoryear{{Neto}, {Gao}, {Bett} \& et al.}{{Neto}
  et~al.}{2007}]{Neto07}
{Neto} A.~F.,  {Gao} L.,  {Bett} P.,    et al. 2007, \mnras, 381, 1450

\bibitem[\protect\citeauthoryear{{Okabe}, {Zhang}, {Finoguenov} \& et
  al.}{{Okabe} et~al.}{2010}]{Okabe10}
{Okabe} N.,  {Zhang} Y.,  {Finoguenov} A.,    et al. 2010, \apj, 721, 875

\bibitem[\protect\citeauthoryear{{Plagge}, {Benson}, {Ade} \& et al.}{{Plagge}
  et~al.}{2010}]{Plagge10}
{Plagge} T.,  {Benson} B.~A.,  {Ade} P.~A.~R.,    et al. 2010, \apj, 716, 1118

\bibitem[\protect\citeauthoryear{{Planck Collaboration}, {Ade}, {Aghanim} \& et
  al.}{{Planck Collaboration} et~al.}{2011a}]{Planck11}
{Planck Collaboration} {Ade} P.~A.~R.,  {Aghanim} N.,    et al. 2011a, ArXiv
  e-prints

\bibitem[\protect\citeauthoryear{{Planck Collaboration}, {Ade}, {Aghanim} \& et
  al.}{{Planck Collaboration} et~al.}{2011b}]{Planck11b}
{Planck Collaboration} {Ade} P.~A.~R.,  {Aghanim} N.,    et al. 2011b, ArXiv
  e-prints

\bibitem[\protect\citeauthoryear{{Planck Collaboration}, {Aghanim}, {Arnaud} \&
  et al.}{{Planck Collaboration} et~al.}{2011c}]{Planck11c}
{Planck Collaboration} {Aghanim} N.,  {Arnaud} M.,    et al. 2011c, ArXiv
  e-prints

\bibitem[\protect\citeauthoryear{{Poole}, {Babul}, {McCarthy} \& et
  al.}{{Poole} et~al.}{2007}]{Poole07}
{Poole} G.~B.,  {Babul} A.,  {McCarthy} I.~G.,    et al. 2007, \mnras, 380, 437

\bibitem[\protect\citeauthoryear{{Poole}, {Fardal}, {Babul} \& et al.}{{Poole}
  et~al.}{2006}]{Poole06}
{Poole} G.~B.,  {Fardal} M.~A.,  {Babul} A.,    et al. 2006, \mnras, 373, 881

\bibitem[\protect\citeauthoryear{{Postman}, {Coe}, {Benitez}, {Bradley} \& et
  al.}{{Postman} et~al.}{2011}]{Postman11}
{Postman} M.,  {Coe} D.,  {Benitez} N.,  {Bradley} L.,    et al. 2011, ArXiv
  e-prints

\bibitem[\protect\citeauthoryear{{Puchwein}, {Sijacki} \&
  {Springel}}{{Puchwein} et~al.}{2008}]{Puchwein08}
{Puchwein} E.,  {Sijacki} D.,    {Springel} V.,  2008, \apjl, 687, L53

\bibitem[\protect\citeauthoryear{{Randall}, {Sarazin} \& {Ricker}}{{Randall}
  et~al.}{2002}]{Randall02}
{Randall} S.~W.,  {Sarazin} C.~L.,    {Ricker} P.~M.,  2002, \apj, 577, 579

\bibitem[\protect\citeauthoryear{{Rasia}, {Ettori}, {Moscardini} \& et
  al.}{{Rasia} et~al.}{2006}]{Rasia06}
{Rasia} E.,  {Ettori} S.,  {Moscardini} L.,    et al. 2006, \mnras, 369, 2013

\bibitem[\protect\citeauthoryear{{Rasia}, {Mazzotta}, {Evrard} \& et
  al.}{{Rasia} et~al.}{2011}]{Rasia11}
{Rasia} E.,  {Mazzotta} P.,  {Evrard} A.,    et al. 2011, \apj, 729, 45

\bibitem[\protect\citeauthoryear{{Rudd} \& {Nagai}}{{Rudd} \&
  {Nagai}}{2009}]{Rudd09}
{Rudd} D.~H.,  {Nagai} D.,  2009, \apjl, 701, L16

\bibitem[\protect\citeauthoryear{{Sayers}, {Golwala}, {Ameglio} \&
  {Pierpaoli}}{{Sayers} et~al.}{2011}]{Sayers11}
{Sayers} J.,  {Golwala} S.~R.,  {Ameglio} S.,    {Pierpaoli} E.,  2011, \apj,
  728, 39

\bibitem[\protect\citeauthoryear{{Sehgal}, {Bode}, {Das},  \& et al.}{{Sehgal}
  et~al.}{2010}]{Sehgal10}
{Sehgal} N.,  {Bode} P.,  {Das} S.,     et al. 2010, \apj, 709, 920

\bibitem[\protect\citeauthoryear{{Sehgal}, {Trac}, {Acquaviva} \& et
  al.}{{Sehgal} et~al.}{2010}]{Sehgal10b}
{Sehgal} N.,  {Trac} H.,  {Acquaviva} V.,    et al. 2010, ArXiv e-prints

\bibitem[\protect\citeauthoryear{{Shaw}, {Holder} \& {Bode}}{{Shaw}
  et~al.}{2008}]{Shaw08}
{Shaw} L.~D.,  {Holder} G.~P.,    {Bode} P.,  2008, \apj, 686, 206

\bibitem[\protect\citeauthoryear{{Shaw}, {Holder} \& {Dudley}}{{Shaw}
  et~al.}{2010}]{Shaw10}
{Shaw} L.~D.,  {Holder} G.~P.,    {Dudley} J.,  2010, \apj, 716, 281

\bibitem[\protect\citeauthoryear{{Sijacki}, {Springel}, {Di Matteo} \&
  {Hernquist}}{{Sijacki} et~al.}{2007}]{Sijacki07}
{Sijacki} D.,  {Springel} V.,  {Di Matteo} T.,    {Hernquist} L.,  2007,
  \mnras, 380, 877

\bibitem[\protect\citeauthoryear{{Smith}, {Edge}, {Eke}, {Nichol}, {Smail} \&
  {Kneib}}{{Smith} et~al.}{2003}]{Smith03}
{Smith} G.~P.,  {Edge} A.~C.,  {Eke} V.~R.,  {Nichol} R.~C.,  {Smail} I.,
  {Kneib} J.-P.,  2003, \apjl, 590, L79

\bibitem[\protect\citeauthoryear{{Springel}}{{Springel}}{2005}]{Springel05}
{Springel} V.,  2005, \mnras, 364, 1105

\bibitem[\protect\citeauthoryear{{Springel} \& {Hernquist}}{{Springel} \&
  {Hernquist}}{2003}]{Springel03}
{Springel} V.,  {Hernquist} L.,  2003, \mnras, 339, 289

\bibitem[\protect\citeauthoryear{{Springel}, {White}, {Jenkins} \& et
  al.}{{Springel} et~al.}{2005}]{Millenium}
{Springel} V.,  {White} S.~D.~M.,  {Jenkins} A.,    et al. 2005, \nat, 435, 629

\bibitem[\protect\citeauthoryear{{Stanek}, {Rasia}, {Evrard} \& et
  al.}{{Stanek} et~al.}{2010}]{Stanek10}
{Stanek} R.,  {Rasia} E.,  {Evrard} A.~E., et al. 2010, \apj, 715, 1508

\bibitem[\protect\citeauthoryear{{Tormen}, {Bouchet} \& {White}}{{Tormen}
  et~al.}{1997}]{Tormen97}
{Tormen} G.,  {Bouchet} F.~R.,    {White} S.~D.~M.,  1997, \mnras, 286, 865

\bibitem[\protect\citeauthoryear{{Vanderlinde}, {Crawford}, {de Haan} \& et
  al.}{{Vanderlinde} et~al.}{2010}]{Vanderlinde10}
{Vanderlinde} K.,  {Crawford} T.~M.,  {de Haan} T.,    et al. 2010, \apj, 722,
  1180

\bibitem[\protect\citeauthoryear{{Ventimiglia}, {Voit}, {Donahue} \&
  {Ameglio}}{{Ventimiglia} et~al.}{2008}]{Ventimiglia08}
{Ventimiglia} D.~A.,  {Voit} G.~M.,  {Donahue} M.,    {Ameglio} S.,  2008,
  \apj, 685, 118

\bibitem[\protect\citeauthoryear{{Wechsler}, {Bullock}, {Primack} \& et
  al.}{{Wechsler} et~al.}{2002}]{Wechsler02}
{Wechsler} R.~H.,  {Bullock} J.~S.,  {Primack} J.~R.,    et al. 2002, \apj,
  568, 52

\bibitem[\protect\citeauthoryear{{Wechsler}, {Zentner}, {Bullock} \& et
  al.}{{Wechsler} et~al.}{2006}]{Wechsler06}
{Wechsler} R.~H.,  {Zentner} A.~R.,  {Bullock} J.~S.,    et al. 2006, \apj,
  652, 71

\bibitem[\protect\citeauthoryear{{White}}{{White}}{2002}]{White02}
{White} M.,  2002, \apjs, 143, 241

\bibitem[\protect\citeauthoryear{{Wik}, {Sarazin}, {Ricker} \& {Randall}}{{Wik}
  et~al.}{2008}]{Wik08}
{Wik} D.~R.,  {Sarazin} C.~L.,  {Ricker} P.~M.,    {Randall} S.~W.,  2008,
  \apj, 680, 17

\bibitem[\protect\citeauthoryear{{Williamson}, {Benson}, {High}, {Vanderlinde}
  \& et al.}{{Williamson} et~al.}{2011}]{Williamson11}
{Williamson} R.,  {Benson} B.~A.,  {High} F.~W.,  {Vanderlinde} K.,    et al.
  2011, ArXiv e-prints

\bibitem[\protect\citeauthoryear{{Wong} \& {Sarazin}}{{Wong} \&
  {Sarazin}}{2009}]{Wong09}
{Wong} K.-W.,  {Sarazin} C.~L.,  2009, \apj, 707, 1141

\bibitem[\protect\citeauthoryear{{Yang}, {Bhattacharya} \& {Ricker}}{{Yang}
  et~al.}{2010}]{Yang10}
{Yang} H.,  {Bhattacharya} S.,    {Ricker} P.~M.,  2010, \apj, 725, 1124

\bibitem[\protect\citeauthoryear{{Yang}, {Ricker} \& {Sutter}}{{Yang}
  et~al.}{2009}]{Yang09}
{Yang} H.,  {Ricker} P.~M.,    {Sutter} P.~M.,  2009, \apj, 699, 315

\bibitem[\protect\citeauthoryear{{Yoshida}, {Colberg}, {White} \& et
  al.}{{Yoshida} et~al.}{2001}]{Yoshida01}
{Yoshida} N.,  {Colberg} J.,  {White} S.~D.~M.,    et al. 2001, \mnras, 325,
  803

\end{thebibliography}

\label{lastpage}
\end{document}